\documentclass[aps,prb,10pt,twocolumn]{revtex4-1}
\usepackage{graphicx}

\begin{document}

\title{Constructing \textit{ab initio} models of ultra-thin Al--AlO$_{\text{x}}$--Al barriers}

\author{Timothy C. DuBois}
\email[]{tim.dubois@tcqp.science}
\affiliation{Chemical and Quantum Physics, School of Applied Sciences, RMIT University, Melbourne, 3001, Australia}

\author{Martin J. Cyster}
\affiliation{Chemical and Quantum Physics, School of Applied Sciences, RMIT University, Melbourne, 3001, Australia}

\author{George Opletal}
\affiliation{Chemical and Quantum Physics, School of Applied Sciences, RMIT University, Melbourne, 3001, Australia}

\author{Salvy P. Russo}
\affiliation{Chemical and Quantum Physics, School of Applied Sciences, RMIT University, Melbourne, 3001, Australia}

\author{Jared H. Cole}
\affiliation{Chemical and Quantum Physics, School of Applied Sciences, RMIT University, Melbourne, 3001, Australia}

\date{\today}

\begin{abstract}
The microscopic structure of ultra-thin oxide barriers often plays a major role in modern nano-electronic devices.  In the case of superconducting electronic circuits, their operation depends on the electrical non-linearity provided by one or more such oxide layers in the form of ultra-thin tunnel barriers (also known as Josephson junctions). Currently available fabrication techniques manufacture an amorphous oxide barrier, which is attributed as a major noise source within the device. The nature of this noise is currently an open question and requires both experimental and theoretical investigation. Here, we present a methodology for constructing atomic scale computational models of Josephson junctions using a combination of molecular mechanics, empirical and ab initio methods. These junctions consist of ultra-thin amorphous aluminium-oxide layers sandwiched between crystalline aluminium. The stability and structure of these barriers as a function of density and stoichiometry are investigated, which we compare to experimentally observed parameters.
\end{abstract}

\maketitle

\section{Introduction}
Josephson junctions: ultra-thin insulating layers sandwiched between layers of superconducting metal, are the fundamental building blocks of the next generation of quantum electronics.
Examples include superconducting quantum-bits~\cite{Clarke1988,Martinis2002,Plourde2005,Deppe2007,Schreier2008}, low power Rapid-Single-Flux-Quantum circuits \cite{Mukhanov1987}, Superconducting Quantum Interference Devices~\cite{Weinstock1996} and non-linear elements for single quanta microwave electronics~\cite{Astafiev2007, Astafiev2010, Wilson2011, Hoi2013, Hoi2013a}.
In all these cases, Josephson junctions provide the non-linear element that allows quantum effects to manifest in the voltage, current or magnetic flux signatures of these circuits.
Recent work on superconducting qubits~\cite{Dutta1981, Shnirman2005} has shown that a key limiting factor in quantum electronics is the existence of loss mechanisms, which can be traced to material defects in the oxide coating (and protecting) the metallic circuits, as well as the oxide which forms the Josephson junction tunnel barrier.
Recent experimental probes of so-called `strongly coupled' defects~\cite{Neeley2008, Lupascu2009, Lisenfeld2010} have shown that they can be individually addressed and manipulated, and mostly likely reside within the junction~\cite{Lacquaniti2012}.
It is therefore fundamentally important to identify and ideally remove these defects as a source of loss and imperfection in quantum circuits.

From a quantum simulation point of view, this is not a trivial problem.
The oxide barrier of a junction is amorphous, so crystalline symmetries cannot be used to reduce the state space.
Added to this is the far more fundamental issue that we currently do not understand what forms the defects of interest.
Various microscopic models exist, including hydrogenic dangling bonds~\cite{Martinis2005,Jameson2011,Holder2013, Gordon2014}, charged surface states~\cite{Choi2009, Lee2014} and delocalisation of the oxygen atoms themselves~\cite{DuBois2013, DuBois2015}.
As well as atomistic models, a range of effective defect state models also exist such as phonon dressing of electronic states~\cite{Agarwal2013}, metal-insulator gap states~\cite{Choi2009} and Andreev bound state models~\cite{DeSousa2009}.
One possible way of distinguishing between these options is to develop complete atomistic models of the Josephson junction and study the configuration of the amorphous layer.
Forming such atomistic models using molecular mechanics and \textit{ab initio} methods is the focus of this paper.

\section{The Josephson Junction Formation Process}

Josephson junctions may be constructed from any superconducting material with any insulating or non-superconducting metal barrier to invoke a weak link coupling.
A popular material choice involves the use of aluminium as the superconducting material, and an amorphous oxide layer as an insulating barrier.

Shadow evaporation is a common technique used to fabricate a system such as this, where two metallic layers are deposited from different angles with an intervening oxidation step.
This is usually performed using a Dolan bridge, which obscures part of the substrate during each metal deposition step \cite{Dolan1977}.
It has more recently been shown that junction fabrication can be performed without the requirement of this bridge \cite{Lecocq2011}.
Regardless of the process chosen, the oxidation of the aluminium does not result in a set of crystalline monolayers, but a non-uniform amorphous layer varying in stoichiometry~\cite{Park2002, Tan2005}, density~\cite{Barbour1998} and thickness~\cite{Gloos2003,Aref2014,Zeng2014} (nominally $\sim\!2$ nm).  Although epitaxial growth of aluminium-oxide barriers has been demonstrated~\cite{Oh2006}, this technique is not yet mainstream as it is considerably more difficult than conventional shadow mask evaporation.
It is therefore the amorphous oxide formation which needs to be investigated predominantly, in order to obtain results from simulation which are applicable to future fabrication work.

Simulating oxide layer growth is in general a difficult problem as the time scale of the oxide growth ($\sim\,$minutes) is many orders of magnitude greater than typically achievable molecular dynamics timescales (ps--ns).
One standard approach is to perform the simulation at elevated temperatures and gas pressures ($\ge$ 1~atm)~\cite{Campbell1999, Zhou2005, Hasnaoui2005}.
This accelerates the oxidation process, making the computation feasible on current high performance computing infrastructure.
However, it also removes the simulation from the reality of experimental junction formation, where pressures range between $10^{-9}$ and $10^{-3}$~atm~\cite{Morohashi1987, Kohlstedt1993, Jeurgens2002}.
It remains to be seen whether any fundamental physics is neglected by adopting this approximation.

An alternative approach is to form an amorphous layer via direct melt and quench~\cite{Vashishta2008,Sheng2012}.
This method has the advantage of computational simplicity and speed, however the resulting layers are not necessarily representative of the true physical situation and therefore benchmarking against other methods and experiment is critical.
Generating stoichiometry or density gradients across an artificial junction is not something that can be simulated directly using this process, so to investigate the effect of these properties, a number of constant density and stoichiometry models were produced. A more sophisticated method, closely mimicking the oxygen deposition process and examining the effects of layer thickness will be considered in future work.

\section{Model construction}\label{sec:model}
\begin{figure}[tbh]
\centering
\includegraphics[width=\columnwidth]{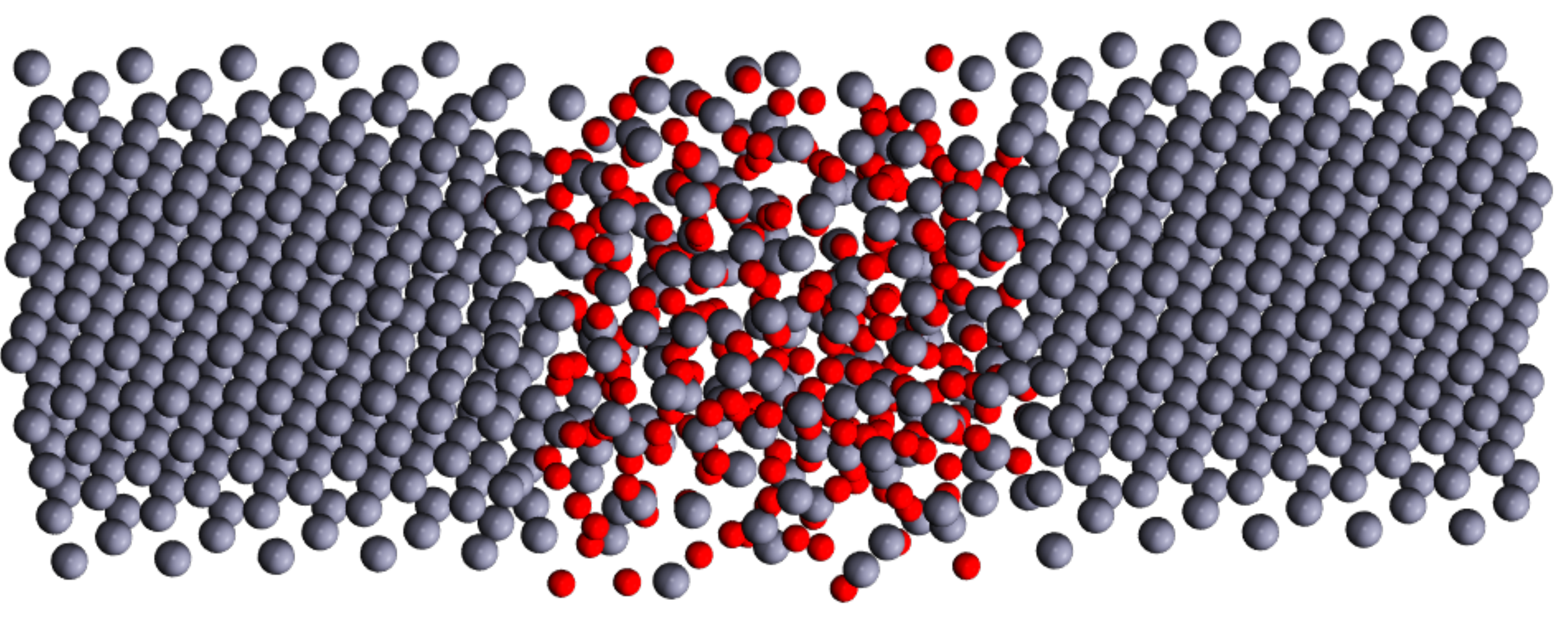}
\caption{\label{fig:povray}Model of a Josephson junction comprised of aluminium (grey) and oxygen (red). Two superconducting regions composed only of aluminium, separated by an amorphous AlO$_{1.25}$ barrier with a density $0.75$ times that of corundum.}%
\end{figure}

To obtain realistic, high precision atomic positions, computational models of the junction were created using a combination of molecular mechanics and Density Functional Theory (DFT).
A $4\!\times\!4\!\times\!5$ supercell of bulk aluminium representing both the top and bottom slabs was relaxed in the DFT code \texttt{VASP}~\cite{Kresse1994, Kresse1996, Kresse1996a} using a projector-augmented wave (PAW) potential~\cite{Kresse1999, Blochl1994}, obtaining a $16.168\times16.168\times20.183$ \AA\ cell.
Exchange-correlation interactions were evaluated using the PBE functional~\cite{Perdew1996}; a $7\!\times\!7\!\times\!7$ $\Gamma$ centered Monkhorst Pack K point mesh and a plane wave cutoff of $250 \; \rm{eV}$.

Formation of the amorphous AlO$_{\rm{x}}$ layers required a number of preparation steps to accurately represent experimental results.
The low temperature and pressure phase of aluminium oxide (commonly referred to as corundum or $\alpha$--Al$_2$O$_3$) was used as a basis for all the constructed junction models.
Experimental investigations of stoichiometry suggest, in general, an oxygen deficiency with oxide O/Al ratios varying between $0.6$ and $1.4$~\cite{Tan2005}, which are highly dependent on the fabrication process.
In response to this, we construct models with four stoichiometries: AlO$_{0.8}$, AlO$_{1.0}$, AlO$_{1.25}$ and AlO$_{1.5}$.
The oxide density may also be an important formation variable.
For simplicity we identify oxide density in multiples of the (average) corundum density: $4.05 \; \rm{g/cm}^3$, and construct junctions with $0.5,\,0.625,\,0.75,\,0.875\,\&\,1.0$ density multiples for each stoichiometry listed above.
A value of $3.2 \; \rm{g/cm}^3$ is typical~\cite{Barbour1998} (which corresponds to a density multiple of 0.8), although theoretical predictions suggest altering the density of this barrier may suppress noise sources of the junction~\cite{DuBois2013}.

Using AlO$_{1.25}$ with a density multiple of $0.75$ as an example, a $6\!\times\!6\!\times\!1$ supercell of corundum was geometry optimised in the software package \texttt{GULP}~\cite{Gale2003}, employing the empirical Streitz-Mintmire potential~\cite{Streitz1994} which can capture the variable oxygen charge states when present in a predominantly metallic environment. This capability is particularly important here, as a Josephson junction has two metal-oxide interfaces.
This large superstructure was required due to the trigonal nature of the lattice, as it was then cut down such that the $xy$ plane of the bulk aluminium slab could be covered.
A non-periodic slab of corundum measuring $16.168 \times 16.168 \times 11.982$ \AA\ was the result of this process.
Oxygen atoms were randomly removed from the corundum lattice until the appropriate stoichiometry of AlO$_{1.25}$ was obtained and the cell was shortened in the $z$-direction to achieve a $0.75$ fractional multiple of the corundum density.
These changes add quite a lot of force onto the structure, so a geometry optimisation (in \texttt{GULP}) was undertaken at this stage to minimise energy contributions.
To simulate the oxygen deposition phase and generate the amorphous nature of these layers, the structure was then annealed using NVT molecular dynamics at $3000$ K with a $1$ fs step size for $3 \; \rm{\mu s}$ and quenched to $350$ K over a $1.5 \; \rm{\mu s}$ period.

The AlO$_{1.25}$ layer was inserted between two bulk Al supercells described above with $0.5$ \AA\ of vacuum space on each side.
The junction was further annealed to simulate a metal--metal--oxide interface reconstruction using \texttt{VASP} NVT Molecular Dynamics at $300$ K until equilibrium was reached (approximately $250$ ionic steps), then geometry optimised using a $2\!\times\!2\!\times\!1$ $\Gamma$ centered Monkhorst Pack K point mesh and a $450$ eV plane wave cutoff to obtain the final model, depicted in Figure~\ref{fig:povray}.

For comparison, junctions were also modelled without the added computational overhead of DFT by solely employing \texttt{GULP} and the Streitz-Mintmire potential.
The construction process of these models matches the procedure above, but interchanges the \textit{ab initio} optimisations of the oxide layer with an empirical framework.

\section{Results and Discussion}

\begin{figure}[tbh]
\centering
\includegraphics[width=\columnwidth]{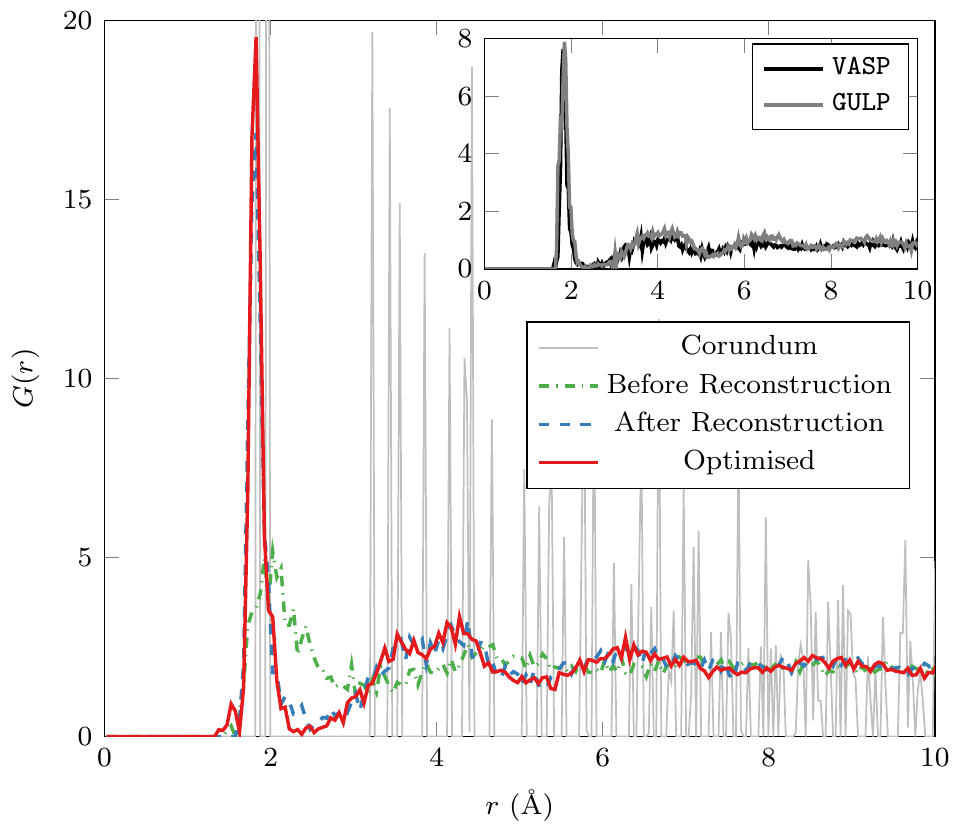}
\caption{\label{fig:groptis}Evolution of the oxygen projected radial distribution function $G(r)$. Crystalline corundum (thin, gray); Metal---metal--oxide interface reconstruction before (dash dotted, green) and after (dashed, blue); final optimised geometry (thick, red).  Inset: Oxygen projected $G(r)$ computed using \textit{ab initio}
 (\texttt{VASP}) and empirical (\texttt{GULP}) methods, showing no statistically significant differences.}%
\end{figure}

To validate our models against experimental observations, we perform a number of statistical tests to scrutinize the structures.
First we must ensure that the oxide layer of the junctions are in fact amorphous in nature.
We employ a projected radial distribution function
\begin{equation}
G(r) = \lim_{dr \to 0}\frac{p(r)}{4\pi\left(N_{\mathrm{pairs}}/V\right)r^2dr}
\end{equation}
where $r$ is the distance between a pair of particles, $p(r)$ is the average number of atom pairs found at a distance between $r$ and $r + dr$, $V$ is the total volume of the system, and $N_{\mathrm{pairs}}$ is the number of unique pairs of atoms~\cite{Levine2011}.
This function was calculated for each stoichiometry and density configuration using oxygen as the reference species, and aluminium atoms in the amorphous region along with the superconducting bulk as the projection species. Figure \ref{fig:groptis} depicts the results of this analysis.

A major peak is visible centered around $1.85$ \AA, which corresponds to convolution of the two Al--O bond distances, $1.852$ \AA\ and $1.971$ \AA\ of the corundum crystal~\cite{Ishizawa1980}.
For a crystalline $G(r)$ this peak is deconvolved to two delta functions (see Figure \ref{fig:groptis} and the discussion below), where here we see a broadening of the statistics and hence differences in neighbour distances: diverging from a crystalline form.
Moving away from this peak to larger distance separations, we see the statistics tending toward a uniform result similar to what a liquid would produce under this analysis.
These two features represent an amorphous system quite well, as close range order suggests a connection to the crystalline form whilst long range order no longer agrees with such periodic conditions.
It's also significant to note that we don't observe neighbours closer than $\sim\!1.5$ \AA\ which is a good indication that the models do not have non-physical neighbour forces acting on atoms.

Most importantly, this trend is almost uniform across all the modeled junctions, which indicate the process outlined in Section \ref{sec:model} is capable of producing amorphous oxides whilst varying other physical parameters of the system.
An evolution of the important steps in the procedure is depicted in Figure \ref{fig:groptis}.

The corundum $G(r)$ (thin, gray) is a complicated structure due to the 30 atom unit cell of the crystal, however it is clear from this figure where much of the amorphous structure originates from.
Specifically the $1.852$ \AA\ and $1.971$ \AA\ Al--O bond distance contributions and the void in the $2$--$3$ \AA\ range.
After the melt/quench phase of the procedure the lattice still appears liquid--like (dash dotted, green).
Whilst the quench cycle minimises the possibility of atoms position very close to one another due to an excess of kinetic energy, it still appears to exhibit liquid behaviour.
This may be a shortcoming of the Streitz-Mintmire potentials ability to capture the relevant physics, however this is rectified after the metal---metal--oxide interface reconstruction is completed (dashed, blue) using the \textit{ab initio} methods.
Finally, the geometry optimisation (thick, red) yields a smoother $1.85$ \AA\ peak and recovers some of the void region around $2$ \AA.

The inset of Figure \ref{fig:groptis} compares the optimal $G(r)$ results for both the \texttt{VASP} and \texttt{GULP} simulations.
Whilst these results are very similar, the \texttt{GULP} simulation actually produces a drastically different final structure.
We find under \texttt{GULP} simulation that stoichiometric ratios higher than 1:1 are not stable and oxygen atoms diffuse into the metallic regions until a stoichiometric ratio of at most 1:1 is achieved.
As a result of this excess oxygen diffusion, the junction width can increase by up to 30\% or more over the course of the simulation.
At high densities and stoichiometries (higher than typical amorphous alumina) some expansion of the oxide region is also seen in the \textit{ab initio} simulations, although this effect is much less pronounced.
Higher oxygen mobility in \texttt{GULP} could be attributed to shortcomings of the empirical potential, but we see very little increase in oxide distribution during the optimisation phase -- suggesting that the details of the Nos\'{e}-Hoover thermostat routine employed during the MD simulation may play a role.

The total energy of a computational model is a good indication of the structure's electronic stability.
Due to the stoichiometry changes invoked in the oxygen depleted models, not all structures have the same number of atoms.
This gives structures with more atoms (such as AlO$_{1.5}$) additional electronegativity which in turn results in a deeper potential well and a large total energy.
In order to be able to validly compare systems of different stoichiometry, we normalise the total energy of each system by a factor $|F|$.
$F = \sum_{k} \mu_k N_k$ is calculated as the linear combination of the number of atoms of chemical species $k$ ($N_k$) by the chemical potential of that species ($\mu_k$), where $k=\{\mathrm{Al}, \mathrm{O}\}$.
The chemical potential for aluminium, $\mu_{\mathrm{Al}}$ was obtained by calculating the DFT total energy of a $4\!\times\!4\!\times\!5$ supercell of bulk Al and dividing by the number of atoms in the supercell. Similarly the chemical potential of oxygen was obtained from calculating the DFT total energy of a $2\!\times\!2\!\times\!2$ supercell of bulk $\mathrm{Al_2O_3}$ using $\mu_{\mathrm{O}} = (\mu_{\mathrm{Al_2O_3}}-2\mu_{\mathrm{Al}})/3$, where $\mu_{\mathrm{Al_2O_3}}$ is the total energy of a molecular unit of $\mathrm{Al_2O_3}$.
The factor $F$ (essentially the free energy at $T=0$) effectively allows one to partition the total energy of the system using the chemical potentials of each component species, as a means to compare the energies of systems with differing number of chemical components.

\begin{figure*}[tbhp]
\centering
\includegraphics[width=\textwidth]{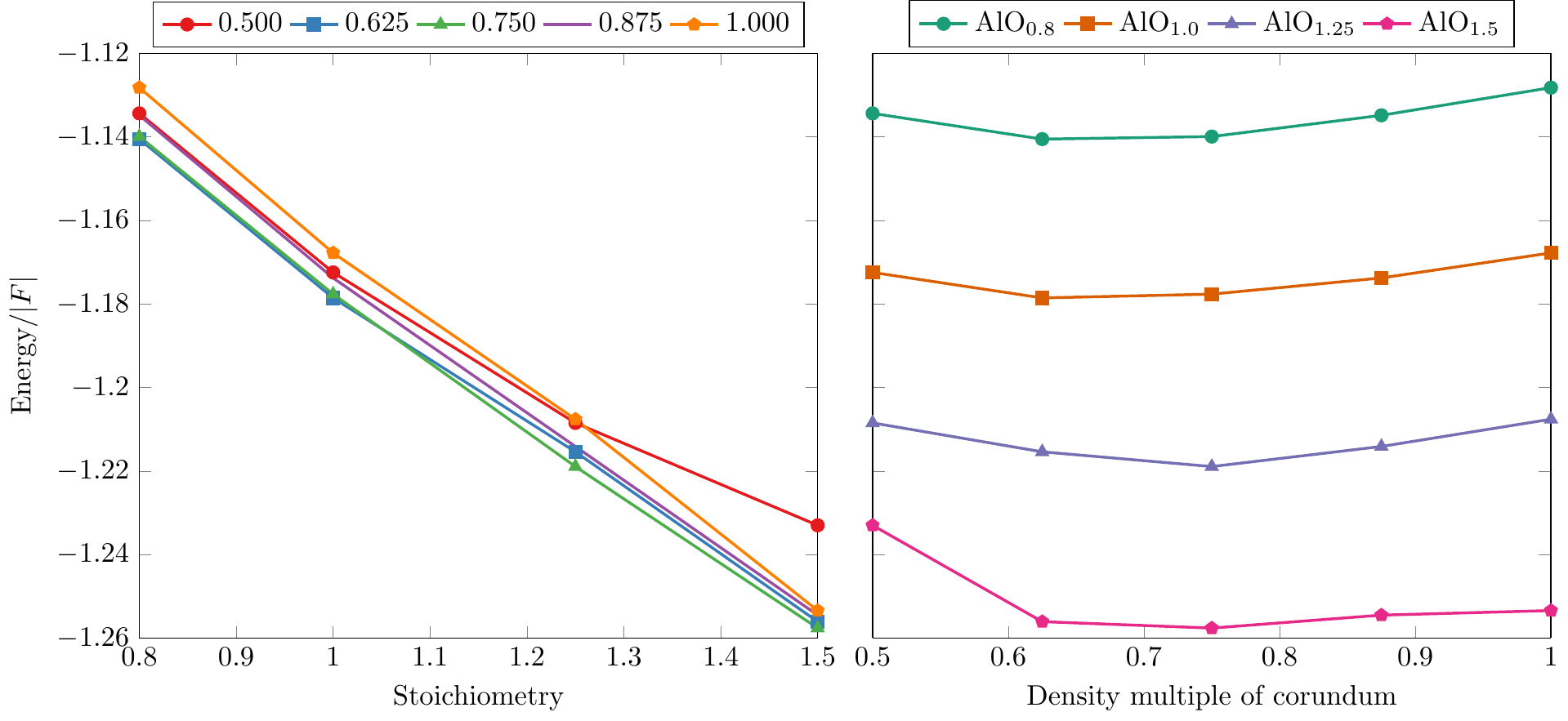}
\caption{\label{fig:energy}Normalised total energy for various junction models with stoichiometry (left) and density (right).  Although the energy is strongly dependent on stoichiometry, we see a trend to optimal densities of approximately 75\% of the density of corundum.}%
\end{figure*}

It is clear from Figure \ref{fig:energy} that stoichiometry plays a larger role in energy minimisation than density, and that the structures would prefer additional oxygen to minimise internal forces.
This suggests that fabrication processes that generate oxygen deficiencies may be inviting the inclusion of alien species or oxygenic site hopping in an attempt to rectify this offset.

Density changes seem to alter the energy contribution marginally.
Minimum energies correspond to density multiples between $0.6$ and $0.75$, slightly lower than typical constructions of $3.2 \; \rm{g/cm}^3$~\cite{Barbour1998} (an 0.8 density multiple); which may indicate another method of experimentally optimising the junction formation process.

Coordination number is a useful metric which allows for some insight into both the crystallinity of the structures being analysed, and their similarity to fabricated junctions. For instance, in the corundum structure every aluminium ion is coordinated with six oxygen ions. In amorphous alumina, the proportion of 6-coordinated aluminium as compared to 4-coordinated aluminium is an experimentally accessible quantity and has been reported on previously~\cite{ElMashri1983}. However, in order to establish this ratio it is assumed that there is a bimodal distribution of octahedral (AlO$_6$) and tetrahedral (AlO$_4$) coordination. Ratios of AlO$_6$:AlO$_4$ are quoted in a range from 80:20 to 30:70, depending on the method by which the oxide layer was formed~\cite{Bourdillon1984}. More modern techniques using nuclear magnetic resonance (NMR) are also able to resolve the AlO$_5$ coordination~\cite{Lee2009}.

\begin{figure*}[tbhp]
\centering
\includegraphics[width=\textwidth]{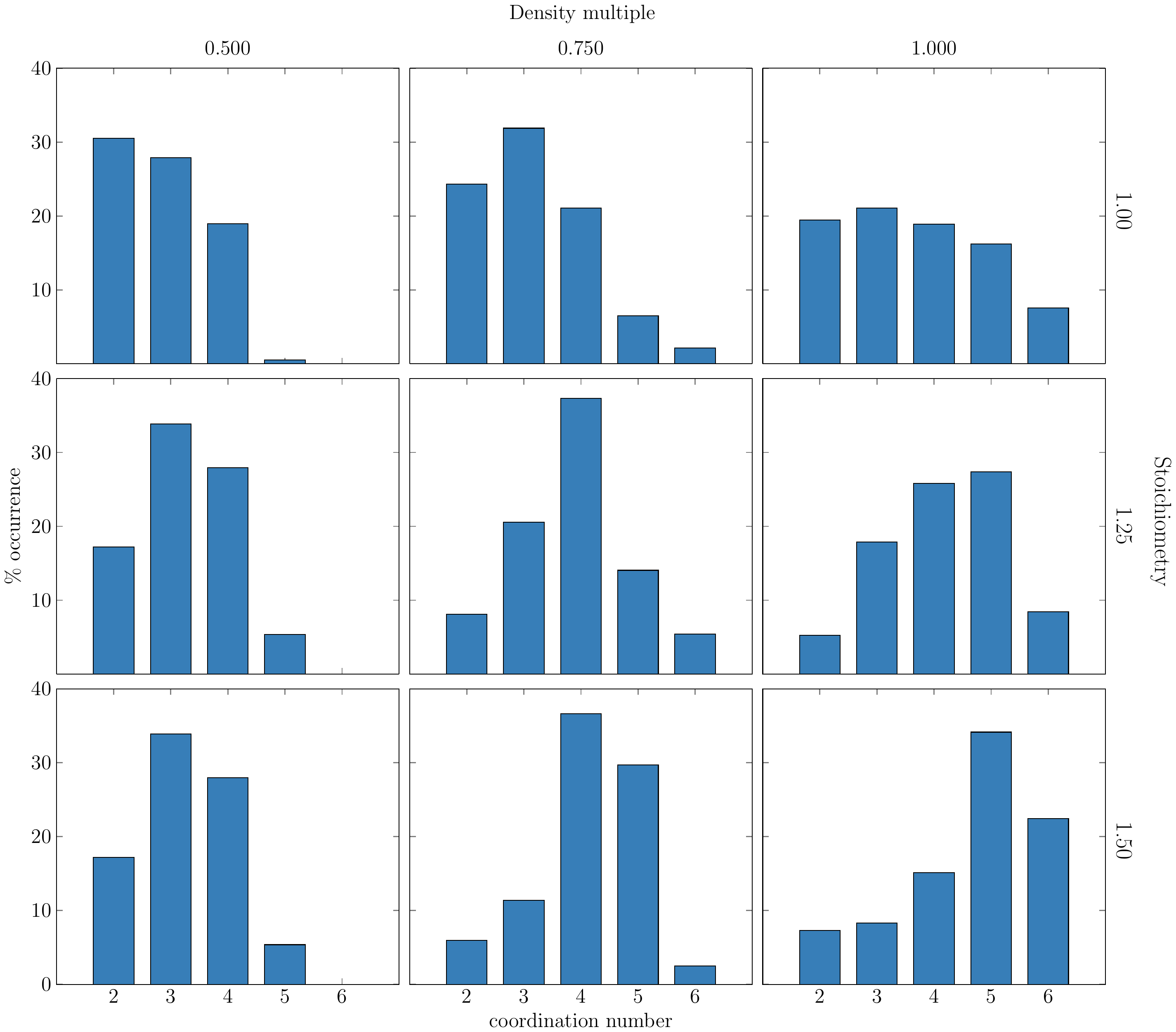}
\caption{\label{fig:coordinationnumber}Distribution of oxygen coordination about aluminium as a function of density and stoichiometry, showing a tendency to higher coordination number with increasing density or stoichiometry.}%
\end{figure*}

Figure~\ref{fig:coordinationnumber} shows the distribution of oxygen coordination about aluminium as a function of density and stoichiometry. These results are calculated using an Al--O bond length cutoff of $2.5$ \AA, which corresponds to the first minimum after the nearest neighbour peak in the $G(r)$ (see Figure \ref{fig:groptis}). As one would expect, the coordination number (for Al--O bonding) increases with increasing density or stoichiometry. We also note that there exists a reasonable proportion of 2- and 3-coordinated aluminium atoms, which persists at high density and stoichiometry. In order to compare directly to previous experimental and theoretical work, we compute the ratio of 4-, 5- and 6-coordination for Al--O bonding, matching the stoichiometry of 1.5 and assuming the density multiple closest to experimental values (0.750). The results are presented in Table~\ref{tab:coord_comp}. We observe excellent agreement, both before and after the \textit{ab initio} optimisation.

\begin{table}[tbh]
\caption{Relative proportions of 4-, 5- and 6-coordinated aluminium atoms within the oxide layer for a density of 0.75 and stoichiometry of 1.5.}
{\begin{tabular}{@{}lccc}\toprule
 & $4$ (\%) & $5$ (\%) & $6$ (\%)\\
\colrule
\texttt{VASP} (before optimisation)	& $57$ & $39$ & $4$ \\
\texttt{VASP} (after optimisation)	& $53$ & $43$ & $4$ \\
Lee \textit{et al.} \cite{Lee2009}; experiment	& $55 \pm 3$ & $42 \pm 3$ & $3 \pm 2$ \\
Momida \textit{et al.} \cite{Momida2011}; theory   & $60.4$ & $29.2$ & $10.4$ \\
\botrule
\end{tabular}}
\label{tab:coord_comp}
\end{table}

\section{Conclusions}

Precise computational models of Josephson junctions are becoming crucial to efforts to reduce dissipation and loss in superconducting circuits.
The limits of computational resources mean that full \textit{ab initio} models are computationally intractable.
However, a combination of \textit{ab initio} and empirical models holds promise for developing flexible and efficient simulation approaches.
We have constructed models of amorphous aluminium-oxide barriers, sandwiched between crystalline aluminium.
Through comparisons with both previous theoretical analysis and experimental measurements, we have shown that the resulting structures are representative of those fabricated experimentally.
The structure of such junctions can be used as input conditions to potential microscopic models for either charge or magnetic defects in Josephson junctions.
Through this approach, free parameters in existing phenomenological defect models can be determined via information directly obtained from the atomic positions.

\section*{Acknowledgements}
This research was supported under the Australian Research Council's Discovery Projects funding scheme (project number DP140100375). Computational resources were provided at the NCI National Facility systems at the Australian National University through the National Computational Merit Allocation Scheme supported by the Australian Government.

\bibliography{junctions}

\begin{thebibliography}{59}%
\makeatletter
\providecommand \@ifxundefined [1]{%
 \@ifx{#1\undefined}
}%
\providecommand \@ifnum [1]{%
 \ifnum #1\expandafter \@firstoftwo
 \else \expandafter \@secondoftwo
 \fi
}%
\providecommand \@ifx [1]{%
 \ifx #1\expandafter \@firstoftwo
 \else \expandafter \@secondoftwo
 \fi
}%
\providecommand \natexlab [1]{#1}%
\providecommand \enquote  [1]{``#1''}%
\providecommand \bibnamefont  [1]{#1}%
\providecommand \bibfnamefont [1]{#1}%
\providecommand \citenamefont [1]{#1}%
\providecommand \href@noop [0]{\@secondoftwo}%
\providecommand \href [0]{\begingroup \@sanitize@url \@href}%
\providecommand \@href[1]{\@@startlink{#1}\@@href}%
\providecommand \@@href[1]{\endgroup#1\@@endlink}%
\providecommand \@sanitize@url [0]{\catcode `\\12\catcode `\$12\catcode
  `\&12\catcode `\#12\catcode `\^12\catcode `\_12\catcode `\%12\relax}%
\providecommand \@@startlink[1]{}%
\providecommand \@@endlink[0]{}%
\providecommand \url  [0]{\begingroup\@sanitize@url \@url }%
\providecommand \@url [1]{\endgroup\@href {#1}{\urlprefix }}%
\providecommand \urlprefix  [0]{URL }%
\providecommand \Eprint [0]{\href }%
\providecommand \doibase [0]{http://dx.doi.org/}%
\providecommand \selectlanguage [0]{\@gobble}%
\providecommand \bibinfo  [0]{\@secondoftwo}%
\providecommand \bibfield  [0]{\@secondoftwo}%
\providecommand \translation [1]{[#1]}%
\providecommand \BibitemOpen [0]{}%
\providecommand \bibitemStop [0]{}%
\providecommand \bibitemNoStop [0]{.\EOS\space}%
\providecommand \EOS [0]{\spacefactor3000\relax}%
\providecommand \BibitemShut  [1]{\csname bibitem#1\endcsname}%
\let\auto@bib@innerbib\@empty
\bibitem [{\citenamefont {Clarke}\ \emph {et~al.}(1988)\citenamefont {Clarke},
  \citenamefont {Cleland}, \citenamefont {Devoret}, \citenamefont {Esteve},\
  and\ \citenamefont {Martinis}}]{Clarke1988}%
  \BibitemOpen
  \bibfield  {author} {\bibinfo {author} {\bibfnamefont {J.}~\bibnamefont
  {Clarke}}, \bibinfo {author} {\bibfnamefont {A.~N.}\ \bibnamefont {Cleland}},
  \bibinfo {author} {\bibfnamefont {M.~H.}\ \bibnamefont {Devoret}}, \bibinfo
  {author} {\bibfnamefont {D.}~\bibnamefont {Esteve}}, \ and\ \bibinfo {author}
  {\bibfnamefont {J.~M.}\ \bibnamefont {Martinis}},\ }\href {\doibase
  10.1126/science.239.4843.992} {\bibfield  {journal} {\bibinfo  {journal}
  {Science}\ }\textbf {\bibinfo {volume} {239}},\ \bibinfo {pages} {992}
  (\bibinfo {year} {1988})}\BibitemShut {NoStop}%
\bibitem [{\citenamefont {Martinis}\ \emph {et~al.}(2002)\citenamefont
  {Martinis}, \citenamefont {Nam}, \citenamefont {Aumentado},\ and\
  \citenamefont {Urbina}}]{Martinis2002}%
  \BibitemOpen
  \bibfield  {author} {\bibinfo {author} {\bibfnamefont {J.}~\bibnamefont
  {Martinis}}, \bibinfo {author} {\bibfnamefont {S.}~\bibnamefont {Nam}},
  \bibinfo {author} {\bibfnamefont {J.}~\bibnamefont {Aumentado}}, \ and\
  \bibinfo {author} {\bibfnamefont {C.}~\bibnamefont {Urbina}},\ }\href
  {\doibase 10.1103/PhysRevLett.89.117901} {\bibfield  {journal} {\bibinfo
  {journal} {Physical Review Letters}\ }\textbf {\bibinfo {volume} {89}},\
  \bibinfo {pages} {117901} (\bibinfo {year} {2002})}\BibitemShut {NoStop}%
\bibitem [{\citenamefont {Plourde}\ \emph {et~al.}(2005)\citenamefont
  {Plourde}, \citenamefont {Robertson}, \citenamefont {Reichardt},
  \citenamefont {Hime}, \citenamefont {Linzen}, \citenamefont {Wu},\ and\
  \citenamefont {Clarke}}]{Plourde2005}%
  \BibitemOpen
  \bibfield  {author} {\bibinfo {author} {\bibfnamefont {B.~L.~T.}\
  \bibnamefont {Plourde}}, \bibinfo {author} {\bibfnamefont {T.~L.}\
  \bibnamefont {Robertson}}, \bibinfo {author} {\bibfnamefont {P.~a.}\
  \bibnamefont {Reichardt}}, \bibinfo {author} {\bibfnamefont {T.}~\bibnamefont
  {Hime}}, \bibinfo {author} {\bibfnamefont {S.}~\bibnamefont {Linzen}},
  \bibinfo {author} {\bibfnamefont {C.~E.}\ \bibnamefont {Wu}}, \ and\ \bibinfo
  {author} {\bibfnamefont {J.}~\bibnamefont {Clarke}},\ }\href {\doibase
  10.1103/PhysRevB.72.060506} {\bibfield  {journal} {\bibinfo  {journal}
  {Physical Review B - Condensed Matter and Materials Physics}\ }\textbf
  {\bibinfo {volume} {72}},\ \bibinfo {pages} {1} (\bibinfo {year} {2005})},\
  \Eprint {http://arxiv.org/abs/0501679} {arXiv:0501679 [cond-mat]}
  \BibitemShut {NoStop}%
\bibitem [{\citenamefont {Deppe}\ \emph {et~al.}(2007)\citenamefont {Deppe},
  \citenamefont {Mariantoni}, \citenamefont {Menzel}, \citenamefont {Saito},
  \citenamefont {Kakuyanagi}, \citenamefont {Tanaka}, \citenamefont {Meno},
  \citenamefont {Semba}, \citenamefont {Takayanagi},\ and\ \citenamefont
  {Gross}}]{Deppe2007}%
  \BibitemOpen
  \bibfield  {author} {\bibinfo {author} {\bibfnamefont {F.}~\bibnamefont
  {Deppe}}, \bibinfo {author} {\bibfnamefont {M.}~\bibnamefont {Mariantoni}},
  \bibinfo {author} {\bibfnamefont {E.~P.}\ \bibnamefont {Menzel}}, \bibinfo
  {author} {\bibfnamefont {S.}~\bibnamefont {Saito}}, \bibinfo {author}
  {\bibfnamefont {K.}~\bibnamefont {Kakuyanagi}}, \bibinfo {author}
  {\bibfnamefont {H.}~\bibnamefont {Tanaka}}, \bibinfo {author} {\bibfnamefont
  {T.}~\bibnamefont {Meno}}, \bibinfo {author} {\bibfnamefont {K.}~\bibnamefont
  {Semba}}, \bibinfo {author} {\bibfnamefont {H.}~\bibnamefont {Takayanagi}}, \
  and\ \bibinfo {author} {\bibfnamefont {R.}~\bibnamefont {Gross}},\ }\href
  {\doibase 10.1103/PhysRevB.76.214503} {\bibfield  {journal} {\bibinfo
  {journal} {Physical Review B - Condensed Matter and Materials Physics}\
  }\textbf {\bibinfo {volume} {76}},\ \bibinfo {pages} {1} (\bibinfo {year}
  {2007})}\BibitemShut {NoStop}%
\bibitem [{\citenamefont {Schreier}\ \emph {et~al.}(2008)\citenamefont
  {Schreier}, \citenamefont {Houck}, \citenamefont {Koch}, \citenamefont
  {Schuster}, \citenamefont {Johnson}, \citenamefont {Chow}, \citenamefont
  {Gambetta}, \citenamefont {Majer}, \citenamefont {Frunzio}, \citenamefont
  {Devoret}, \citenamefont {Girvin},\ and\ \citenamefont
  {Schoelkopf}}]{Schreier2008}%
  \BibitemOpen
  \bibfield  {author} {\bibinfo {author} {\bibfnamefont {J.}~\bibnamefont
  {Schreier}}, \bibinfo {author} {\bibfnamefont {A.}~\bibnamefont {Houck}},
  \bibinfo {author} {\bibfnamefont {J.}~\bibnamefont {Koch}}, \bibinfo {author}
  {\bibfnamefont {D.}~\bibnamefont {Schuster}}, \bibinfo {author}
  {\bibfnamefont {B.}~\bibnamefont {Johnson}}, \bibinfo {author} {\bibfnamefont
  {J.}~\bibnamefont {Chow}}, \bibinfo {author} {\bibfnamefont {J.}~\bibnamefont
  {Gambetta}}, \bibinfo {author} {\bibfnamefont {J.}~\bibnamefont {Majer}},
  \bibinfo {author} {\bibfnamefont {L.}~\bibnamefont {Frunzio}}, \bibinfo
  {author} {\bibfnamefont {M.}~\bibnamefont {Devoret}}, \bibinfo {author}
  {\bibfnamefont {S.}~\bibnamefont {Girvin}}, \ and\ \bibinfo {author}
  {\bibfnamefont {R.}~\bibnamefont {Schoelkopf}},\ }\href {\doibase
  10.1103/PhysRevB.77.180502} {\bibfield  {journal} {\bibinfo  {journal}
  {Physical Review B}\ }\textbf {\bibinfo {volume} {77}},\ \bibinfo {pages}
  {180502} (\bibinfo {year} {2008})}\BibitemShut {NoStop}%
\bibitem [{\citenamefont {Mukhanov}\ \emph {et~al.}(1987)\citenamefont
  {Mukhanov}, \citenamefont {Semenov},\ and\ \citenamefont
  {Likharev}}]{Mukhanov1987}%
  \BibitemOpen
  \bibfield  {author} {\bibinfo {author} {\bibfnamefont {O.}~\bibnamefont
  {Mukhanov}}, \bibinfo {author} {\bibfnamefont {V.}~\bibnamefont {Semenov}}, \
  and\ \bibinfo {author} {\bibfnamefont {K.}~\bibnamefont {Likharev}},\ }\href
  {\doibase 10.1109/TMAG.1987.1064951} {\enquote {\bibinfo {title} {{Ultimate
  performance of the RSFQ logic circuits}},}\ } (\bibinfo {year}
  {1987})\BibitemShut {NoStop}%
\bibitem [{\citenamefont {Clarke}(1996)}]{Weinstock1996}%
  \BibitemOpen
  \bibfield  {author} {\bibinfo {author} {\bibfnamefont {J.}~\bibnamefont
  {Clarke}},\ }in\ \href {\doibase 10.1007/978-94-011-5674-5} {\emph {\bibinfo
  {booktitle} {SQUID Sensors: Fundamentals, Fabrication and Applications}}},\
  \bibinfo {editor} {edited by\ \bibinfo {editor} {\bibfnamefont
  {H.}~\bibnamefont {Weinstock}}}\ (\bibinfo  {publisher} {Springer},\ \bibinfo
  {year} {1996})\ Chap.~\bibinfo {chapter} {1}, pp.\ \bibinfo {pages}
  {1--62}\BibitemShut {NoStop}%
\bibitem [{\citenamefont {Astafiev}\ \emph {et~al.}(2007)\citenamefont
  {Astafiev}, \citenamefont {Inomata}, \citenamefont {Niskanen}, \citenamefont
  {Yamamoto}, \citenamefont {Pashkin}, \citenamefont {Nakamura},\ and\
  \citenamefont {Tsai}}]{Astafiev2007}%
  \BibitemOpen
  \bibfield  {author} {\bibinfo {author} {\bibfnamefont {O.}~\bibnamefont
  {Astafiev}}, \bibinfo {author} {\bibfnamefont {K.}~\bibnamefont {Inomata}},
  \bibinfo {author} {\bibfnamefont {a.~O.}\ \bibnamefont {Niskanen}}, \bibinfo
  {author} {\bibfnamefont {T.}~\bibnamefont {Yamamoto}}, \bibinfo {author}
  {\bibfnamefont {Y.~a.}\ \bibnamefont {Pashkin}}, \bibinfo {author}
  {\bibfnamefont {Y.}~\bibnamefont {Nakamura}}, \ and\ \bibinfo {author}
  {\bibfnamefont {J.~S.}\ \bibnamefont {Tsai}},\ }\href {\doibase
  10.1038/nature06141} {\bibfield  {journal} {\bibinfo  {journal} {Nature}\
  }\textbf {\bibinfo {volume} {449}},\ \bibinfo {pages} {588} (\bibinfo {year}
  {2007})},\ \Eprint {http://arxiv.org/abs/0710.0936} {arXiv:0710.0936}
  \BibitemShut {NoStop}%
\bibitem [{\citenamefont {Astafiev}\ \emph {et~al.}(2010)\citenamefont
  {Astafiev}, \citenamefont {Zagoskin}, \citenamefont {Abdumalikov},
  \citenamefont {Pashkin}, \citenamefont {Yamamoto}, \citenamefont {Inomata},
  \citenamefont {Nakamura},\ and\ \citenamefont {Tsai}}]{Astafiev2010}%
  \BibitemOpen
  \bibfield  {author} {\bibinfo {author} {\bibfnamefont {O.}~\bibnamefont
  {Astafiev}}, \bibinfo {author} {\bibfnamefont {a.~M.}\ \bibnamefont
  {Zagoskin}}, \bibinfo {author} {\bibfnamefont {a.~a.}\ \bibnamefont
  {Abdumalikov}}, \bibinfo {author} {\bibfnamefont {Y.~a.}\ \bibnamefont
  {Pashkin}}, \bibinfo {author} {\bibfnamefont {T.}~\bibnamefont {Yamamoto}},
  \bibinfo {author} {\bibfnamefont {K.}~\bibnamefont {Inomata}}, \bibinfo
  {author} {\bibfnamefont {Y.}~\bibnamefont {Nakamura}}, \ and\ \bibinfo
  {author} {\bibfnamefont {J.~S.}\ \bibnamefont {Tsai}},\ }\href {\doibase
  10.1126/science.1181918} {\bibfield  {journal} {\bibinfo  {journal} {Science
  (New York, N.Y.)}\ }\textbf {\bibinfo {volume} {327}},\ \bibinfo {pages}
  {840} (\bibinfo {year} {2010})},\ \Eprint {http://arxiv.org/abs/1002.4944}
  {arXiv:1002.4944} \BibitemShut {NoStop}%
\bibitem [{\citenamefont {Wilson}\ \emph {et~al.}(2011)\citenamefont {Wilson},
  \citenamefont {Johansson}, \citenamefont {Pourkabirian}, \citenamefont
  {Johansson}, \citenamefont {Duty}, \citenamefont {Nori},\ and\ \citenamefont
  {Delsing}}]{Wilson2011}%
  \BibitemOpen
  \bibfield  {author} {\bibinfo {author} {\bibfnamefont {C.~M.}\ \bibnamefont
  {Wilson}}, \bibinfo {author} {\bibfnamefont {G.}~\bibnamefont {Johansson}},
  \bibinfo {author} {\bibfnamefont {a.}~\bibnamefont {Pourkabirian}}, \bibinfo
  {author} {\bibfnamefont {J.~R.}\ \bibnamefont {Johansson}}, \bibinfo {author}
  {\bibfnamefont {T.}~\bibnamefont {Duty}}, \bibinfo {author} {\bibfnamefont
  {F.}~\bibnamefont {Nori}}, \ and\ \bibinfo {author} {\bibfnamefont
  {P.}~\bibnamefont {Delsing}},\ }\href {\doibase 10.1038/nature10561}
  {\bibfield  {journal} {\bibinfo  {journal} {Nature}\ }\textbf {\bibinfo
  {volume} {479}},\ \bibinfo {pages} {12} (\bibinfo {year} {2011})},\ \Eprint
  {http://arxiv.org/abs/1105.4714} {arXiv:1105.4714} \BibitemShut {NoStop}%
\bibitem [{\citenamefont {Hoi}\ \emph {et~al.}(2013{\natexlab{a}})\citenamefont
  {Hoi}, \citenamefont {Kockum}, \citenamefont {Palomaki}, \citenamefont
  {Stace}, \citenamefont {Fan}, \citenamefont {Tornberg}, \citenamefont
  {Sathyamoorthy}, \citenamefont {Johansson}, \citenamefont {Delsing},\ and\
  \citenamefont {Wilson}}]{Hoi2013}%
  \BibitemOpen
  \bibfield  {author} {\bibinfo {author} {\bibfnamefont {I.~C.}\ \bibnamefont
  {Hoi}}, \bibinfo {author} {\bibfnamefont {A.~F.}\ \bibnamefont {Kockum}},
  \bibinfo {author} {\bibfnamefont {T.}~\bibnamefont {Palomaki}}, \bibinfo
  {author} {\bibfnamefont {T.~M.}\ \bibnamefont {Stace}}, \bibinfo {author}
  {\bibfnamefont {B.}~\bibnamefont {Fan}}, \bibinfo {author} {\bibfnamefont
  {L.}~\bibnamefont {Tornberg}}, \bibinfo {author} {\bibfnamefont {S.~R.}\
  \bibnamefont {Sathyamoorthy}}, \bibinfo {author} {\bibfnamefont
  {G.}~\bibnamefont {Johansson}}, \bibinfo {author} {\bibfnamefont
  {P.}~\bibnamefont {Delsing}}, \ and\ \bibinfo {author} {\bibfnamefont
  {C.~M.}\ \bibnamefont {Wilson}},\ }\href {\doibase
  10.1103/PhysRevLett.111.053601} {\bibfield  {journal} {\bibinfo  {journal}
  {Physical Review Letters}\ }\textbf {\bibinfo {volume} {111}},\ \bibinfo
  {pages} {2} (\bibinfo {year} {2013}{\natexlab{a}})},\ \Eprint
  {http://arxiv.org/abs/1207.1203} {arXiv:1207.1203} \BibitemShut {NoStop}%
\bibitem [{\citenamefont {Hoi}\ \emph {et~al.}(2013{\natexlab{b}})\citenamefont
  {Hoi}, \citenamefont {Wilson}, \citenamefont {Johansson}, \citenamefont
  {Lindkvist}, \citenamefont {Peropadre}, \citenamefont {Palomaki},\ and\
  \citenamefont {Delsing}}]{Hoi2013a}%
  \BibitemOpen
  \bibfield  {author} {\bibinfo {author} {\bibfnamefont {I.~C.}\ \bibnamefont
  {Hoi}}, \bibinfo {author} {\bibfnamefont {C.~M.}\ \bibnamefont {Wilson}},
  \bibinfo {author} {\bibfnamefont {G.}~\bibnamefont {Johansson}}, \bibinfo
  {author} {\bibfnamefont {J.}~\bibnamefont {Lindkvist}}, \bibinfo {author}
  {\bibfnamefont {B.}~\bibnamefont {Peropadre}}, \bibinfo {author}
  {\bibfnamefont {T.}~\bibnamefont {Palomaki}}, \ and\ \bibinfo {author}
  {\bibfnamefont {P.}~\bibnamefont {Delsing}},\ }\href {\doibase
  10.1088/1367-2630/15/2/025011} {\bibfield  {journal} {\bibinfo  {journal}
  {New Journal of Physics}\ }\textbf {\bibinfo {volume} {15}} (\bibinfo {year}
  {2013}{\natexlab{b}}),\ 10.1088/1367-2630/15/2/025011},\ \Eprint
  {http://arxiv.org/abs/arXiv:1210.4303v1} {arXiv:arXiv:1210.4303v1}
  \BibitemShut {NoStop}%
\bibitem [{\citenamefont {Dutta}\ and\ \citenamefont {Horn}(1981)}]{Dutta1981}%
  \BibitemOpen
  \bibfield  {author} {\bibinfo {author} {\bibfnamefont {P.}~\bibnamefont
  {Dutta}}\ and\ \bibinfo {author} {\bibfnamefont {P.}~\bibnamefont {Horn}},\
  }\href {\doibase 10.1103/RevModPhys.53.497} {\bibfield  {journal} {\bibinfo
  {journal} {Reviews of Modern Physics}\ }\textbf {\bibinfo {volume} {53}},\
  \bibinfo {pages} {497} (\bibinfo {year} {1981})}\BibitemShut {NoStop}%
\bibitem [{\citenamefont {Shnirman}\ \emph {et~al.}(2005)\citenamefont
  {Shnirman}, \citenamefont {Sch\"{o}n}, \citenamefont {Martin},\ and\
  \citenamefont {Makhlin}}]{Shnirman2005}%
  \BibitemOpen
  \bibfield  {author} {\bibinfo {author} {\bibfnamefont {A.}~\bibnamefont
  {Shnirman}}, \bibinfo {author} {\bibfnamefont {G.}~\bibnamefont {Sch\"{o}n}},
  \bibinfo {author} {\bibfnamefont {I.}~\bibnamefont {Martin}}, \ and\ \bibinfo
  {author} {\bibfnamefont {Y.}~\bibnamefont {Makhlin}},\ }\href {\doibase
  10.1103/PhysRevLett.94.127002} {\bibfield  {journal} {\bibinfo  {journal}
  {Physical Review Letters}\ }\textbf {\bibinfo {volume} {94}},\ \bibinfo
  {pages} {127002} (\bibinfo {year} {2005})}\BibitemShut {NoStop}%
\bibitem [{\citenamefont {Neeley}\ \emph {et~al.}(2008)\citenamefont {Neeley},
  \citenamefont {Ansmann}, \citenamefont {Bialczak}, \citenamefont {Hofheinz},
  \citenamefont {Katz}, \citenamefont {Lucero}, \citenamefont {O'Connell},
  \citenamefont {Wang}, \citenamefont {Cleland},\ and\ \citenamefont
  {Martinis}}]{Neeley2008}%
  \BibitemOpen
  \bibfield  {author} {\bibinfo {author} {\bibfnamefont {M.}~\bibnamefont
  {Neeley}}, \bibinfo {author} {\bibfnamefont {M.}~\bibnamefont {Ansmann}},
  \bibinfo {author} {\bibfnamefont {R.~C.}\ \bibnamefont {Bialczak}}, \bibinfo
  {author} {\bibfnamefont {M.}~\bibnamefont {Hofheinz}}, \bibinfo {author}
  {\bibfnamefont {N.}~\bibnamefont {Katz}}, \bibinfo {author} {\bibfnamefont
  {E.}~\bibnamefont {Lucero}}, \bibinfo {author} {\bibfnamefont
  {A.}~\bibnamefont {O'Connell}}, \bibinfo {author} {\bibfnamefont
  {H.}~\bibnamefont {Wang}}, \bibinfo {author} {\bibfnamefont {A.~N.}\
  \bibnamefont {Cleland}}, \ and\ \bibinfo {author} {\bibfnamefont {J.~M.}\
  \bibnamefont {Martinis}},\ }\href {\doibase 10.1038/nphys972} {\bibfield
  {journal} {\bibinfo  {journal} {Nature Physics}\ }\textbf {\bibinfo {volume}
  {4}},\ \bibinfo {pages} {523} (\bibinfo {year} {2008})}\BibitemShut {NoStop}%
\bibitem [{\citenamefont {Lupaşcu}\ \emph {et~al.}(2009)\citenamefont
  {Lupaşcu}, \citenamefont {Bertet}, \citenamefont {Driessen}, \citenamefont
  {Harmans},\ and\ \citenamefont {Mooij}}]{Lupascu2009}%
  \BibitemOpen
  \bibfield  {author} {\bibinfo {author} {\bibfnamefont {A.}~\bibnamefont
  {Lupaşcu}}, \bibinfo {author} {\bibfnamefont {P.}~\bibnamefont {Bertet}},
  \bibinfo {author} {\bibfnamefont {E.~F.~C.}\ \bibnamefont {Driessen}},
  \bibinfo {author} {\bibfnamefont {C.~J. P.~M.}\ \bibnamefont {Harmans}}, \
  and\ \bibinfo {author} {\bibfnamefont {J.~E.}\ \bibnamefont {Mooij}},\ }\href
  {\doibase 10.1103/PhysRevB.80.172506} {\bibfield  {journal} {\bibinfo
  {journal} {Physical Review B}\ }\textbf {\bibinfo {volume} {80}},\ \bibinfo
  {pages} {172506} (\bibinfo {year} {2009})}\BibitemShut {NoStop}%
\bibitem [{\citenamefont {Lisenfeld}\ \emph {et~al.}(2010)\citenamefont
  {Lisenfeld}, \citenamefont {M\"{u}ller}, \citenamefont {Cole}, \citenamefont
  {Bushev}, \citenamefont {Lukashenko}, \citenamefont {Shnirman},\ and\
  \citenamefont {Ustinov}}]{Lisenfeld2010}%
  \BibitemOpen
  \bibfield  {author} {\bibinfo {author} {\bibfnamefont {J.}~\bibnamefont
  {Lisenfeld}}, \bibinfo {author} {\bibfnamefont {C.}~\bibnamefont
  {M\"{u}ller}}, \bibinfo {author} {\bibfnamefont {J.~H.}\ \bibnamefont
  {Cole}}, \bibinfo {author} {\bibfnamefont {P.}~\bibnamefont {Bushev}},
  \bibinfo {author} {\bibfnamefont {A.}~\bibnamefont {Lukashenko}}, \bibinfo
  {author} {\bibfnamefont {A.}~\bibnamefont {Shnirman}}, \ and\ \bibinfo
  {author} {\bibfnamefont {A.~V.}\ \bibnamefont {Ustinov}},\ }\href {\doibase
  10.1103/PhysRevLett.105.230504} {\bibfield  {journal} {\bibinfo  {journal}
  {Physical Review Letters}\ }\textbf {\bibinfo {volume} {105}},\ \bibinfo
  {pages} {230504} (\bibinfo {year} {2010})}\BibitemShut {NoStop}%
\bibitem [{\citenamefont {Lacquaniti}\ \emph {et~al.}(2012)\citenamefont
  {Lacquaniti}, \citenamefont {Belogolovskii}, \citenamefont {Cassiago},
  \citenamefont {{De Leo}}, \citenamefont {Fretto},\ and\ \citenamefont
  {Sosso}}]{Lacquaniti2012}%
  \BibitemOpen
  \bibfield  {author} {\bibinfo {author} {\bibfnamefont {V.}~\bibnamefont
  {Lacquaniti}}, \bibinfo {author} {\bibfnamefont {M.}~\bibnamefont
  {Belogolovskii}}, \bibinfo {author} {\bibfnamefont {C.}~\bibnamefont
  {Cassiago}}, \bibinfo {author} {\bibfnamefont {N.}~\bibnamefont {{De Leo}}},
  \bibinfo {author} {\bibfnamefont {M.}~\bibnamefont {Fretto}}, \ and\ \bibinfo
  {author} {\bibfnamefont {A.}~\bibnamefont {Sosso}},\ }\href {\doibase
  10.1088/1367-2630/14/2/023025} {\bibfield  {journal} {\bibinfo  {journal}
  {New Journal of Physics}\ }\textbf {\bibinfo {volume} {14}},\ \bibinfo
  {pages} {023025} (\bibinfo {year} {2012})}\BibitemShut {NoStop}%
\bibitem [{\citenamefont {Martinis}\ \emph {et~al.}(2005)\citenamefont
  {Martinis}, \citenamefont {Cooper}, \citenamefont {McDermott}, \citenamefont
  {Steffen}, \citenamefont {Ansmann}, \citenamefont {Osborn}, \citenamefont
  {Cicak}, \citenamefont {Oh}, \citenamefont {Pappas}, \citenamefont
  {Simmonds},\ and\ \citenamefont {Yu}}]{Martinis2005}%
  \BibitemOpen
  \bibfield  {author} {\bibinfo {author} {\bibfnamefont {J.}~\bibnamefont
  {Martinis}}, \bibinfo {author} {\bibfnamefont {K.}~\bibnamefont {Cooper}},
  \bibinfo {author} {\bibfnamefont {R.}~\bibnamefont {McDermott}}, \bibinfo
  {author} {\bibfnamefont {M.}~\bibnamefont {Steffen}}, \bibinfo {author}
  {\bibfnamefont {M.}~\bibnamefont {Ansmann}}, \bibinfo {author} {\bibfnamefont
  {K.}~\bibnamefont {Osborn}}, \bibinfo {author} {\bibfnamefont
  {K.}~\bibnamefont {Cicak}}, \bibinfo {author} {\bibfnamefont
  {S.}~\bibnamefont {Oh}}, \bibinfo {author} {\bibfnamefont {D.}~\bibnamefont
  {Pappas}}, \bibinfo {author} {\bibfnamefont {R.}~\bibnamefont {Simmonds}}, \
  and\ \bibinfo {author} {\bibfnamefont {C.}~\bibnamefont {Yu}},\ }\href
  {\doibase 10.1103/PhysRevLett.95.210503} {\bibfield  {journal} {\bibinfo
  {journal} {Physical Review Letters}\ }\textbf {\bibinfo {volume} {95}},\
  \bibinfo {pages} {210503} (\bibinfo {year} {2005})}\BibitemShut {NoStop}%
\bibitem [{\citenamefont {Jameson}\ \emph {et~al.}(2011)\citenamefont
  {Jameson}, \citenamefont {Ngo}, \citenamefont {Benko}, \citenamefont
  {McVittie}, \citenamefont {Nishi},\ and\ \citenamefont
  {Young}}]{Jameson2011}%
  \BibitemOpen
  \bibfield  {author} {\bibinfo {author} {\bibfnamefont {J.~R.}\ \bibnamefont
  {Jameson}}, \bibinfo {author} {\bibfnamefont {D.}~\bibnamefont {Ngo}},
  \bibinfo {author} {\bibfnamefont {C.}~\bibnamefont {Benko}}, \bibinfo
  {author} {\bibfnamefont {J.}~\bibnamefont {McVittie}}, \bibinfo {author}
  {\bibfnamefont {Y.}~\bibnamefont {Nishi}}, \ and\ \bibinfo {author}
  {\bibfnamefont {B.}~\bibnamefont {Young}},\ }\href {\doibase
  10.1016/j.jnoncrysol.2011.02.054} {\bibfield  {journal} {\bibinfo  {journal}
  {Journal of Non-Crystalline Solids}\ }\textbf {\bibinfo {volume} {357}},\
  \bibinfo {pages} {2148} (\bibinfo {year} {2011})}\BibitemShut {NoStop}%
\bibitem [{\citenamefont {Holder}\ \emph {et~al.}(2013)\citenamefont {Holder},
  \citenamefont {Osborn}, \citenamefont {Lobb},\ and\ \citenamefont
  {Musgrave}}]{Holder2013}%
  \BibitemOpen
  \bibfield  {author} {\bibinfo {author} {\bibfnamefont {A.~M.}\ \bibnamefont
  {Holder}}, \bibinfo {author} {\bibfnamefont {K.~D.}\ \bibnamefont {Osborn}},
  \bibinfo {author} {\bibfnamefont {C.~J.}\ \bibnamefont {Lobb}}, \ and\
  \bibinfo {author} {\bibfnamefont {C.~B.}\ \bibnamefont {Musgrave}},\ }\href
  {\doibase 10.1103/PhysRevLett.111.065901} {\bibfield  {journal} {\bibinfo
  {journal} {Physical Review Letters}\ }\textbf {\bibinfo {volume} {111}},\
  \bibinfo {pages} {065901} (\bibinfo {year} {2013})}\BibitemShut {NoStop}%
\bibitem [{\citenamefont {Gordon}\ \emph {et~al.}(2014)\citenamefont {Gordon},
  \citenamefont {Abu-farsakh}, \citenamefont {Janotti},\ and\ \citenamefont
  {Walle}}]{Gordon2014}%
  \BibitemOpen
  \bibfield  {author} {\bibinfo {author} {\bibfnamefont {L.}~\bibnamefont
  {Gordon}}, \bibinfo {author} {\bibfnamefont {H.}~\bibnamefont {Abu-farsakh}},
  \bibinfo {author} {\bibfnamefont {A.}~\bibnamefont {Janotti}}, \ and\
  \bibinfo {author} {\bibfnamefont {C.~G. V.~D.}\ \bibnamefont {Walle}},\
  }\href {\doibase 10.1038/srep07590} {\bibfield  {journal} {\bibinfo
  {journal} {Scientific Reports}\ }\textbf {\bibinfo {volume} {4}},\ \bibinfo
  {pages} {7590} (\bibinfo {year} {2014})}\BibitemShut {NoStop}%
\bibitem [{\citenamefont {Choi}\ \emph {et~al.}(2009)\citenamefont {Choi},
  \citenamefont {Lee}, \citenamefont {Louie},\ and\ \citenamefont
  {Clarke}}]{Choi2009}%
  \BibitemOpen
  \bibfield  {author} {\bibinfo {author} {\bibfnamefont {S.}~\bibnamefont
  {Choi}}, \bibinfo {author} {\bibfnamefont {D.-H.}\ \bibnamefont {Lee}},
  \bibinfo {author} {\bibfnamefont {S.}~\bibnamefont {Louie}}, \ and\ \bibinfo
  {author} {\bibfnamefont {J.}~\bibnamefont {Clarke}},\ }\href {\doibase
  10.1103/PhysRevLett.103.197001} {\bibfield  {journal} {\bibinfo  {journal}
  {Physical Review Letters}\ }\textbf {\bibinfo {volume} {103}},\ \bibinfo
  {pages} {197001} (\bibinfo {year} {2009})}\BibitemShut {NoStop}%
\bibitem [{\citenamefont {Lee}\ \emph {et~al.}(2014)\citenamefont {Lee},
  \citenamefont {DuBois},\ and\ \citenamefont {Lordi}}]{Lee2014}%
  \BibitemOpen
  \bibfield  {author} {\bibinfo {author} {\bibfnamefont {D.}~\bibnamefont
  {Lee}}, \bibinfo {author} {\bibfnamefont {J.}~\bibnamefont {DuBois}}, \ and\
  \bibinfo {author} {\bibfnamefont {V.}~\bibnamefont {Lordi}},\ }\href
  {\doibase 10.1103/PhysRevLett.112.017001} {\bibfield  {journal} {\bibinfo
  {journal} {Physical Review Letters}\ }\textbf {\bibinfo {volume} {112}},\
  \bibinfo {pages} {017001} (\bibinfo {year} {2014})}\BibitemShut {NoStop}%
\bibitem [{\citenamefont {DuBois}\ \emph {et~al.}(2013)\citenamefont {DuBois},
  \citenamefont {Per}, \citenamefont {Russo},\ and\ \citenamefont
  {Cole}}]{DuBois2013}%
  \BibitemOpen
  \bibfield  {author} {\bibinfo {author} {\bibfnamefont {T.~C.}\ \bibnamefont
  {DuBois}}, \bibinfo {author} {\bibfnamefont {M.~C.}\ \bibnamefont {Per}},
  \bibinfo {author} {\bibfnamefont {S.~P.}\ \bibnamefont {Russo}}, \ and\
  \bibinfo {author} {\bibfnamefont {J.~H.}\ \bibnamefont {Cole}},\ }\href
  {\doibase 10.1103/PhysRevLett.110.077002} {\bibfield  {journal} {\bibinfo
  {journal} {Physical Review Letters}\ }\textbf {\bibinfo {volume} {110}},\
  \bibinfo {pages} {077002} (\bibinfo {year} {2013})}\BibitemShut {NoStop}%
\bibitem [{\citenamefont {DuBois}\ \emph {et~al.}(2015)\citenamefont {DuBois},
  \citenamefont {Russo},\ and\ \citenamefont {Cole}}]{DuBois2015}%
  \BibitemOpen
  \bibfield  {author} {\bibinfo {author} {\bibfnamefont {T.~C.}\ \bibnamefont
  {DuBois}}, \bibinfo {author} {\bibfnamefont {S.~P.}\ \bibnamefont {Russo}}, \
  and\ \bibinfo {author} {\bibfnamefont {J.~H.}\ \bibnamefont {Cole}},\ }\href
  {\doibase 10.1088/1367-2630/17/2/023017} {\bibfield  {journal} {\bibinfo
  {journal} {New Journal of Physics}\ }\textbf {\bibinfo {volume} {17}},\
  \bibinfo {pages} {13} (\bibinfo {year} {2015})},\ \Eprint
  {http://arxiv.org/abs/1408.5687} {arXiv:1408.5687} \BibitemShut {NoStop}%
\bibitem [{\citenamefont {Agarwal}\ \emph {et~al.}(2013)\citenamefont
  {Agarwal}, \citenamefont {Martin}, \citenamefont {Lukin},\ and\ \citenamefont
  {Demler}}]{Agarwal2013}%
  \BibitemOpen
  \bibfield  {author} {\bibinfo {author} {\bibfnamefont {K.}~\bibnamefont
  {Agarwal}}, \bibinfo {author} {\bibfnamefont {I.}~\bibnamefont {Martin}},
  \bibinfo {author} {\bibfnamefont {M.}~\bibnamefont {Lukin}}, \ and\ \bibinfo
  {author} {\bibfnamefont {E.}~\bibnamefont {Demler}},\ }\href {\doibase
  10.1103/PhysRevB.87.144201} {\bibfield  {journal} {\bibinfo  {journal}
  {Physical Review B}\ }\textbf {\bibinfo {volume} {87}},\ \bibinfo {pages}
  {144201} (\bibinfo {year} {2013})}\BibitemShut {NoStop}%
\bibitem [{\citenamefont {de~Sousa}\ \emph {et~al.}(2009)\citenamefont
  {de~Sousa}, \citenamefont {Whaley}, \citenamefont {Hecht}, \citenamefont {von
  Delft},\ and\ \citenamefont {Wilhelm}}]{DeSousa2009}%
  \BibitemOpen
  \bibfield  {author} {\bibinfo {author} {\bibfnamefont {R.}~\bibnamefont
  {de~Sousa}}, \bibinfo {author} {\bibfnamefont {K.}~\bibnamefont {Whaley}},
  \bibinfo {author} {\bibfnamefont {T.}~\bibnamefont {Hecht}}, \bibinfo
  {author} {\bibfnamefont {J.}~\bibnamefont {von Delft}}, \ and\ \bibinfo
  {author} {\bibfnamefont {F.}~\bibnamefont {Wilhelm}},\ }\href {\doibase
  10.1103/PhysRevB.80.094515} {\bibfield  {journal} {\bibinfo  {journal}
  {Physical Review B}\ }\textbf {\bibinfo {volume} {80}},\ \bibinfo {pages}
  {094515} (\bibinfo {year} {2009})}\BibitemShut {NoStop}%
\bibitem [{\citenamefont {Dolan}(1977)}]{Dolan1977}%
  \BibitemOpen
  \bibfield  {author} {\bibinfo {author} {\bibfnamefont {G.~J.}\ \bibnamefont
  {Dolan}},\ }\href {\doibase 10.1063/1.89690} {\bibfield  {journal} {\bibinfo
  {journal} {Applied Physics Letters}\ }\textbf {\bibinfo {volume} {31}},\
  \bibinfo {pages} {337} (\bibinfo {year} {1977})}\BibitemShut {NoStop}%
\bibitem [{\citenamefont {Lecocq}\ \emph {et~al.}(2011)\citenamefont {Lecocq},
  \citenamefont {Pop}, \citenamefont {Peng}, \citenamefont {Matei},
  \citenamefont {Crozes}, \citenamefont {Fournier}, \citenamefont {Naud},
  \citenamefont {Guichard},\ and\ \citenamefont {Buisson}}]{Lecocq2011}%
  \BibitemOpen
  \bibfield  {author} {\bibinfo {author} {\bibfnamefont {F.}~\bibnamefont
  {Lecocq}}, \bibinfo {author} {\bibfnamefont {I.~M.}\ \bibnamefont {Pop}},
  \bibinfo {author} {\bibfnamefont {Z.}~\bibnamefont {Peng}}, \bibinfo {author}
  {\bibfnamefont {I.}~\bibnamefont {Matei}}, \bibinfo {author} {\bibfnamefont
  {T.}~\bibnamefont {Crozes}}, \bibinfo {author} {\bibfnamefont
  {T.}~\bibnamefont {Fournier}}, \bibinfo {author} {\bibfnamefont
  {C.}~\bibnamefont {Naud}}, \bibinfo {author} {\bibfnamefont {W.}~\bibnamefont
  {Guichard}}, \ and\ \bibinfo {author} {\bibfnamefont {O.}~\bibnamefont
  {Buisson}},\ }\href {\doibase 10.1088/0957-4484/22/31/315302} {\bibfield
  {journal} {\bibinfo  {journal} {Nanotechnology}\ }\textbf {\bibinfo {volume}
  {22}},\ \bibinfo {pages} {315302} (\bibinfo {year} {2011})}\BibitemShut
  {NoStop}%
\bibitem [{\citenamefont {Park}\ \emph {et~al.}(2002)\citenamefont {Park},
  \citenamefont {Bae},\ and\ \citenamefont {Lee}}]{Park2002}%
  \BibitemOpen
  \bibfield  {author} {\bibinfo {author} {\bibfnamefont {B.~G.}\ \bibnamefont
  {Park}}, \bibinfo {author} {\bibfnamefont {J.~Y.}\ \bibnamefont {Bae}}, \
  and\ \bibinfo {author} {\bibfnamefont {T.~D.}\ \bibnamefont {Lee}},\ }\href
  {\doibase 10.1063/1.1447210} {\bibfield  {journal} {\bibinfo  {journal}
  {Journal of Applied Physics}\ }\textbf {\bibinfo {volume} {91}},\ \bibinfo
  {pages} {8789} (\bibinfo {year} {2002})}\BibitemShut {NoStop}%
\bibitem [{\citenamefont {Tan}\ \emph {et~al.}(2005)\citenamefont {Tan},
  \citenamefont {Mather}, \citenamefont {Perrella}, \citenamefont {Read},\ and\
  \citenamefont {Buhrman}}]{Tan2005}%
  \BibitemOpen
  \bibfield  {author} {\bibinfo {author} {\bibfnamefont {E.}~\bibnamefont
  {Tan}}, \bibinfo {author} {\bibfnamefont {P.}~\bibnamefont {Mather}},
  \bibinfo {author} {\bibfnamefont {A.}~\bibnamefont {Perrella}}, \bibinfo
  {author} {\bibfnamefont {J.}~\bibnamefont {Read}}, \ and\ \bibinfo {author}
  {\bibfnamefont {R.}~\bibnamefont {Buhrman}},\ }\href {\doibase
  10.1103/PhysRevB.71.161401} {\bibfield  {journal} {\bibinfo  {journal}
  {Physical Review B}\ }\textbf {\bibinfo {volume} {71}},\ \bibinfo {pages}
  {161401} (\bibinfo {year} {2005})}\BibitemShut {NoStop}%
\bibitem [{\citenamefont {Barbour}\ \emph {et~al.}(1998)\citenamefont
  {Barbour}, \citenamefont {Copeland}, \citenamefont {Dunn}, \citenamefont
  {Missert}, \citenamefont {Montes}, \citenamefont {Son},\ and\ \citenamefont
  {Sullivan}}]{Barbour1998}%
  \BibitemOpen
  \bibfield  {author} {\bibinfo {author} {\bibfnamefont {J.~C.}\ \bibnamefont
  {Barbour}}, \bibinfo {author} {\bibfnamefont {R.~G.}\ \bibnamefont
  {Copeland}}, \bibinfo {author} {\bibfnamefont {R.~G.}\ \bibnamefont {Dunn}},
  \bibinfo {author} {\bibfnamefont {N.}~\bibnamefont {Missert}}, \bibinfo
  {author} {\bibfnamefont {L.~P.}\ \bibnamefont {Montes}}, \bibinfo {author}
  {\bibfnamefont {K.-A.}\ \bibnamefont {Son}}, \ and\ \bibinfo {author}
  {\bibfnamefont {J.~P.}\ \bibnamefont {Sullivan}},\ }in\ \href
  {http://www.osti.gov/scitech/servlets/purl/1916} {\emph {\bibinfo {booktitle}
  {The Electrochemical Society Meeting}}}\ (\bibinfo {year} {1998})\BibitemShut
  {NoStop}%
\bibitem [{\citenamefont {Gloos}\ \emph {et~al.}(2003)\citenamefont {Gloos},
  \citenamefont {Koppinen},\ and\ \citenamefont {Pekola}}]{Gloos2003}%
  \BibitemOpen
  \bibfield  {author} {\bibinfo {author} {\bibfnamefont {K.}~\bibnamefont
  {Gloos}}, \bibinfo {author} {\bibfnamefont {P.~J.}\ \bibnamefont {Koppinen}},
  \ and\ \bibinfo {author} {\bibfnamefont {J.~P.}\ \bibnamefont {Pekola}},\
  }\href {\doibase 10.1088/0953-8984/15/10/320} {\bibfield  {journal} {\bibinfo
   {journal} {Journal of Physics: Condensed Matter}\ }\textbf {\bibinfo
  {volume} {15}},\ \bibinfo {pages} {1733} (\bibinfo {year}
  {2003})}\BibitemShut {NoStop}%
\bibitem [{\citenamefont {Aref}\ \emph {et~al.}(2014)\citenamefont {Aref},
  \citenamefont {Averin}, \citenamefont {van Dijken}, \citenamefont {Ferring},
  \citenamefont {Koberidze}, \citenamefont {Maisi}, \citenamefont {Nguyen},
  \citenamefont {Nieminen}, \citenamefont {Pekola},\ and\ \citenamefont
  {Yao}}]{Aref2014}%
  \BibitemOpen
  \bibfield  {author} {\bibinfo {author} {\bibfnamefont {T.}~\bibnamefont
  {Aref}}, \bibinfo {author} {\bibfnamefont {A.}~\bibnamefont {Averin}},
  \bibinfo {author} {\bibfnamefont {S.}~\bibnamefont {van Dijken}}, \bibinfo
  {author} {\bibfnamefont {A.}~\bibnamefont {Ferring}}, \bibinfo {author}
  {\bibfnamefont {M.}~\bibnamefont {Koberidze}}, \bibinfo {author}
  {\bibfnamefont {V.~F.}\ \bibnamefont {Maisi}}, \bibinfo {author}
  {\bibfnamefont {H.}~\bibnamefont {Nguyen}}, \bibinfo {author} {\bibfnamefont
  {R.~M.}\ \bibnamefont {Nieminen}}, \bibinfo {author} {\bibfnamefont {J.~P.}\
  \bibnamefont {Pekola}}, \ and\ \bibinfo {author} {\bibfnamefont {L.~D.}\
  \bibnamefont {Yao}},\ }\href {\doibase 10.1063/1.4893473} {\bibfield
  {journal} {\bibinfo  {journal} {Journal of Applied Physics}\ }\textbf
  {\bibinfo {volume} {116}},\ \bibinfo {pages} {4} (\bibinfo {year}
  {2014})}\BibitemShut {NoStop}%
\bibitem [{\citenamefont {Zeng}\ \emph {et~al.}(2014)\citenamefont {Zeng},
  \citenamefont {Nik}, \citenamefont {Greibe}, \citenamefont {Wilson},
  \citenamefont {Delsing},\ and\ \citenamefont {Olsson}}]{Zeng2014}%
  \BibitemOpen
  \bibfield  {author} {\bibinfo {author} {\bibfnamefont {L.~J.}\ \bibnamefont
  {Zeng}}, \bibinfo {author} {\bibfnamefont {S.}~\bibnamefont {Nik}}, \bibinfo
  {author} {\bibfnamefont {T.}~\bibnamefont {Greibe}}, \bibinfo {author}
  {\bibfnamefont {C.~M.}\ \bibnamefont {Wilson}}, \bibinfo {author}
  {\bibfnamefont {P.}~\bibnamefont {Delsing}}, \ and\ \bibinfo {author}
  {\bibfnamefont {E.}~\bibnamefont {Olsson}},\ }\href
  {http://arxiv.org/abs/1407.0173} {\bibfield  {journal} {\bibinfo  {journal}
  {arXiv}\ ,\ \bibinfo {pages} {5}} (\bibinfo {year} {2014})},\ \Eprint
  {http://arxiv.org/abs/1407.0173} {arXiv:1407.0173} \BibitemShut {NoStop}%
\bibitem [{\citenamefont {Oh}\ \emph {et~al.}(2006)\citenamefont {Oh},
  \citenamefont {Cicak}, \citenamefont {Kline}, \citenamefont
  {Sillanp\"{a}\"{a}}, \citenamefont {Osborn}, \citenamefont {Whittaker},
  \citenamefont {Simmonds},\ and\ \citenamefont {Pappas}}]{Oh2006}%
  \BibitemOpen
  \bibfield  {author} {\bibinfo {author} {\bibfnamefont {S.}~\bibnamefont
  {Oh}}, \bibinfo {author} {\bibfnamefont {K.}~\bibnamefont {Cicak}}, \bibinfo
  {author} {\bibfnamefont {J.~S.}\ \bibnamefont {Kline}}, \bibinfo {author}
  {\bibfnamefont {M.~a.}\ \bibnamefont {Sillanp\"{a}\"{a}}}, \bibinfo {author}
  {\bibfnamefont {K.~D.}\ \bibnamefont {Osborn}}, \bibinfo {author}
  {\bibfnamefont {J.~D.}\ \bibnamefont {Whittaker}}, \bibinfo {author}
  {\bibfnamefont {R.~W.}\ \bibnamefont {Simmonds}}, \ and\ \bibinfo {author}
  {\bibfnamefont {D.~P.}\ \bibnamefont {Pappas}},\ }\href {\doibase
  10.1103/PhysRevB.74.100502} {\bibfield  {journal} {\bibinfo  {journal}
  {Physical Review B - Condensed Matter and Materials Physics}\ }\textbf
  {\bibinfo {volume} {74}},\ \bibinfo {pages} {1} (\bibinfo {year}
  {2006})}\BibitemShut {NoStop}%
\bibitem [{\citenamefont {Campbell}\ \emph {et~al.}(1999)\citenamefont
  {Campbell}, \citenamefont {Kalia}, \citenamefont {Nakano}, \citenamefont
  {Vashishta}, \citenamefont {Ogata},\ and\ \citenamefont
  {Rodgers}}]{Campbell1999}%
  \BibitemOpen
  \bibfield  {author} {\bibinfo {author} {\bibfnamefont {T.}~\bibnamefont
  {Campbell}}, \bibinfo {author} {\bibfnamefont {R.}~\bibnamefont {Kalia}},
  \bibinfo {author} {\bibfnamefont {A.}~\bibnamefont {Nakano}}, \bibinfo
  {author} {\bibfnamefont {P.}~\bibnamefont {Vashishta}}, \bibinfo {author}
  {\bibfnamefont {S.}~\bibnamefont {Ogata}}, \ and\ \bibinfo {author}
  {\bibfnamefont {S.}~\bibnamefont {Rodgers}},\ }\href {\doibase
  10.1103/PhysRevLett.82.4866} {\bibfield  {journal} {\bibinfo  {journal}
  {Physical Review Letters}\ }\textbf {\bibinfo {volume} {82}},\ \bibinfo
  {pages} {4866} (\bibinfo {year} {1999})}\BibitemShut {NoStop}%
\bibitem [{\citenamefont {Zhou}\ and\ \citenamefont {Wadley}(2005)}]{Zhou2005}%
  \BibitemOpen
  \bibfield  {author} {\bibinfo {author} {\bibfnamefont {X.}~\bibnamefont
  {Zhou}}\ and\ \bibinfo {author} {\bibfnamefont {H.}~\bibnamefont {Wadley}},\
  }\href {\doibase 10.1103/PhysRevB.71.054418} {\bibfield  {journal} {\bibinfo
  {journal} {Physical Review B}\ }\textbf {\bibinfo {volume} {71}},\ \bibinfo
  {pages} {054418} (\bibinfo {year} {2005})}\BibitemShut {NoStop}%
\bibitem [{\citenamefont {Hasnaoui}\ \emph {et~al.}(2005)\citenamefont
  {Hasnaoui}, \citenamefont {Politano}, \citenamefont {Salazar}, \citenamefont
  {Aral}, \citenamefont {Kalia}, \citenamefont {Nakano},\ and\ \citenamefont
  {Vashishta}}]{Hasnaoui2005}%
  \BibitemOpen
  \bibfield  {author} {\bibinfo {author} {\bibfnamefont {A.}~\bibnamefont
  {Hasnaoui}}, \bibinfo {author} {\bibfnamefont {O.}~\bibnamefont {Politano}},
  \bibinfo {author} {\bibfnamefont {J.}~\bibnamefont {Salazar}}, \bibinfo
  {author} {\bibfnamefont {G.}~\bibnamefont {Aral}}, \bibinfo {author}
  {\bibfnamefont {R.}~\bibnamefont {Kalia}}, \bibinfo {author} {\bibfnamefont
  {A.}~\bibnamefont {Nakano}}, \ and\ \bibinfo {author} {\bibfnamefont
  {P.}~\bibnamefont {Vashishta}},\ }\href {\doibase 10.1016/j.susc.2005.01.043}
  {\bibfield  {journal} {\bibinfo  {journal} {Surface Science}\ }\textbf
  {\bibinfo {volume} {579}},\ \bibinfo {pages} {47} (\bibinfo {year}
  {2005})}\BibitemShut {NoStop}%
\bibitem [{\citenamefont {Morohashi}\ and\ \citenamefont
  {Hasuo}(1987)}]{Morohashi1987}%
  \BibitemOpen
  \bibfield  {author} {\bibinfo {author} {\bibfnamefont {S.}~\bibnamefont
  {Morohashi}}\ and\ \bibinfo {author} {\bibfnamefont {S.}~\bibnamefont
  {Hasuo}},\ }\href {\doibase 10.1063/1.338348} {\bibfield  {journal} {\bibinfo
   {journal} {Journal of Applied Physics}\ }\textbf {\bibinfo {volume} {61}},\
  \bibinfo {pages} {4835} (\bibinfo {year} {1987})}\BibitemShut {NoStop}%
\bibitem [{\citenamefont {Kohlstedt}\ \emph {et~al.}(1993)\citenamefont
  {Kohlstedt}, \citenamefont {Hallmanns}, \citenamefont {Nevirkovets},
  \citenamefont {Guggi},\ and\ \citenamefont {Heiden}}]{Kohlstedt1993}%
  \BibitemOpen
  \bibfield  {author} {\bibinfo {author} {\bibfnamefont {H.}~\bibnamefont
  {Kohlstedt}}, \bibinfo {author} {\bibfnamefont {G.}~\bibnamefont
  {Hallmanns}}, \bibinfo {author} {\bibfnamefont {I.}~\bibnamefont
  {Nevirkovets}}, \bibinfo {author} {\bibfnamefont {D.}~\bibnamefont {Guggi}},
  \ and\ \bibinfo {author} {\bibfnamefont {C.}~\bibnamefont {Heiden}},\ }\href
  {http://ieeexplore.ieee.org/xpls/abs\_all.jsp?arnumber=233939} {\bibfield
  {journal} {\bibinfo  {journal} {IEEE Transactions on Applied
  Superconductivity}\ }\textbf {\bibinfo {volume} {3}},\ \bibinfo {pages}
  {2197} (\bibinfo {year} {1993})}\BibitemShut {NoStop}%
\bibitem [{\citenamefont {Jeurgens}\ \emph {et~al.}(2002)\citenamefont
  {Jeurgens}, \citenamefont {Sloof}, \citenamefont {Tichelaar},\ and\
  \citenamefont {Mittemeijer}}]{Jeurgens2002}%
  \BibitemOpen
  \bibfield  {author} {\bibinfo {author} {\bibfnamefont {L.}~\bibnamefont
  {Jeurgens}}, \bibinfo {author} {\bibfnamefont {W.}~\bibnamefont {Sloof}},
  \bibinfo {author} {\bibfnamefont {F.}~\bibnamefont {Tichelaar}}, \ and\
  \bibinfo {author} {\bibfnamefont {E.}~\bibnamefont {Mittemeijer}},\ }\href
  {\doibase 10.1016/S0040-6090(02)00787-3} {\bibfield  {journal} {\bibinfo
  {journal} {Thin Solid Films}\ }\textbf {\bibinfo {volume} {418}},\ \bibinfo
  {pages} {89} (\bibinfo {year} {2002})}\BibitemShut {NoStop}%
\bibitem [{\citenamefont {Vashishta}\ \emph {et~al.}(2008)\citenamefont
  {Vashishta}, \citenamefont {Kalia}, \citenamefont {Nakano},\ and\
  \citenamefont {Rino}}]{Vashishta2008}%
  \BibitemOpen
  \bibfield  {author} {\bibinfo {author} {\bibfnamefont {P.}~\bibnamefont
  {Vashishta}}, \bibinfo {author} {\bibfnamefont {R.~K.}\ \bibnamefont
  {Kalia}}, \bibinfo {author} {\bibfnamefont {A.}~\bibnamefont {Nakano}}, \
  and\ \bibinfo {author} {\bibfnamefont {J.~P.}\ \bibnamefont {Rino}},\ }\href
  {\doibase 10.1063/1.2901171} {\bibfield  {journal} {\bibinfo  {journal}
  {Journal of Applied Physics}\ }\textbf {\bibinfo {volume} {103}},\ \bibinfo
  {pages} {083504} (\bibinfo {year} {2008})}\BibitemShut {NoStop}%
\bibitem [{\citenamefont {Sheng}\ \emph {et~al.}(2012)\citenamefont {Sheng},
  \citenamefont {Ma},\ and\ \citenamefont {Kramer}}]{Sheng2012}%
  \BibitemOpen
  \bibfield  {author} {\bibinfo {author} {\bibfnamefont {H.~W.}\ \bibnamefont
  {Sheng}}, \bibinfo {author} {\bibfnamefont {E.}~\bibnamefont {Ma}}, \ and\
  \bibinfo {author} {\bibfnamefont {M.~J.}\ \bibnamefont {Kramer}},\ }\href
  {\doibase 10.1007/s11837-012-0360-y} {\bibfield  {journal} {\bibinfo
  {journal} {Jom}\ }\textbf {\bibinfo {volume} {64}},\ \bibinfo {pages} {856}
  (\bibinfo {year} {2012})}\BibitemShut {NoStop}%
\bibitem [{\citenamefont {Kresse}\ and\ \citenamefont
  {Hafner}(1994)}]{Kresse1994}%
  \BibitemOpen
  \bibfield  {author} {\bibinfo {author} {\bibfnamefont {G.}~\bibnamefont
  {Kresse}}\ and\ \bibinfo {author} {\bibfnamefont {J.}~\bibnamefont
  {Hafner}},\ }\href {\doibase 10.1088/0953-8984/6/40/015} {\bibfield
  {journal} {\bibinfo  {journal} {Journal of Physics: Condensed Matter}\
  }\textbf {\bibinfo {volume} {6}},\ \bibinfo {pages} {8245} (\bibinfo {year}
  {1994})}\BibitemShut {NoStop}%
\bibitem [{\citenamefont {Kresse}\ and\ \citenamefont
  {Furthm\"{u}ller}(1996)}]{Kresse1996}%
  \BibitemOpen
  \bibfield  {author} {\bibinfo {author} {\bibfnamefont {G.}~\bibnamefont
  {Kresse}}\ and\ \bibinfo {author} {\bibfnamefont {J.}~\bibnamefont
  {Furthm\"{u}ller}},\ }\href {\doibase 10.1016/0927-0256(96)00008-0}
  {\bibfield  {journal} {\bibinfo  {journal} {Computational Materials Science}\
  }\textbf {\bibinfo {volume} {6}},\ \bibinfo {pages} {15} (\bibinfo {year}
  {1996})}\BibitemShut {NoStop}%
\bibitem [{\citenamefont {Kresse}(1996)}]{Kresse1996a}%
  \BibitemOpen
  \bibfield  {author} {\bibinfo {author} {\bibfnamefont {G.}~\bibnamefont
  {Kresse}},\ }\href {\doibase 10.1103/PhysRevB.54.11169} {\bibfield  {journal}
  {\bibinfo  {journal} {Physical Review B}\ }\textbf {\bibinfo {volume} {54}},\
  \bibinfo {pages} {11169} (\bibinfo {year} {1996})}\BibitemShut {NoStop}%
\bibitem [{\citenamefont {Kresse}(1999)}]{Kresse1999}%
  \BibitemOpen
  \bibfield  {author} {\bibinfo {author} {\bibfnamefont {G.}~\bibnamefont
  {Kresse}},\ }\href {\doibase 10.1103/PhysRevB.59.1758} {\bibfield  {journal}
  {\bibinfo  {journal} {Physical Review B}\ }\textbf {\bibinfo {volume} {59}},\
  \bibinfo {pages} {1758} (\bibinfo {year} {1999})}\BibitemShut {NoStop}%
\bibitem [{\citenamefont {Bl\"{o}chl}(1994)}]{Blochl1994}%
  \BibitemOpen
  \bibfield  {author} {\bibinfo {author} {\bibfnamefont {P.~E.}\ \bibnamefont
  {Bl\"{o}chl}},\ }\href {\doibase 10.1103/PhysRevB.50.17953} {\bibfield
  {journal} {\bibinfo  {journal} {Physical Review B}\ }\textbf {\bibinfo
  {volume} {50}},\ \bibinfo {pages} {17953} (\bibinfo {year}
  {1994})}\BibitemShut {NoStop}%
\bibitem [{\citenamefont {Perdew}\ \emph {et~al.}(1996)\citenamefont {Perdew},
  \citenamefont {Burke},\ and\ \citenamefont {Ernzerhof}}]{Perdew1996}%
  \BibitemOpen
  \bibfield  {author} {\bibinfo {author} {\bibfnamefont {J.~P.}\ \bibnamefont
  {Perdew}}, \bibinfo {author} {\bibfnamefont {K.}~\bibnamefont {Burke}}, \
  and\ \bibinfo {author} {\bibfnamefont {M.}~\bibnamefont {Ernzerhof}},\ }\href
  {\doibase 10.1103/PhysRevLett.77.3865} {\bibfield  {journal} {\bibinfo
  {journal} {Physical Review Letters}\ }\textbf {\bibinfo {volume} {77}},\
  \bibinfo {pages} {3865} (\bibinfo {year} {1996})}\BibitemShut {NoStop}%
\bibitem [{\citenamefont {Gale}\ and\ \citenamefont {Rohl}(2003)}]{Gale2003}%
  \BibitemOpen
  \bibfield  {author} {\bibinfo {author} {\bibfnamefont {J.~D.}\ \bibnamefont
  {Gale}}\ and\ \bibinfo {author} {\bibfnamefont {A.~L.}\ \bibnamefont
  {Rohl}},\ }\href {\doibase 10.1080/0892702031000104887} {\bibfield  {journal}
  {\bibinfo  {journal} {Molecular Simulation}\ }\textbf {\bibinfo {volume}
  {29}},\ \bibinfo {pages} {291} (\bibinfo {year} {2003})}\BibitemShut
  {NoStop}%
\bibitem [{\citenamefont {Streitz}\ and\ \citenamefont
  {Mintmire}(1994)}]{Streitz1994}%
  \BibitemOpen
  \bibfield  {author} {\bibinfo {author} {\bibfnamefont {F.}~\bibnamefont
  {Streitz}}\ and\ \bibinfo {author} {\bibfnamefont {J.}~\bibnamefont
  {Mintmire}},\ }\href {\doibase 10.1103/PhysRevB.50.11996} {\bibfield
  {journal} {\bibinfo  {journal} {Physical Review B}\ }\textbf {\bibinfo
  {volume} {50}},\ \bibinfo {pages} {11996} (\bibinfo {year}
  {1994})}\BibitemShut {NoStop}%
\bibitem [{\citenamefont {Levine}\ \emph {et~al.}(2011)\citenamefont {Levine},
  \citenamefont {Stone},\ and\ \citenamefont {Kohlmeyer}}]{Levine2011}%
  \BibitemOpen
  \bibfield  {author} {\bibinfo {author} {\bibfnamefont {B.~G.}\ \bibnamefont
  {Levine}}, \bibinfo {author} {\bibfnamefont {J.~E.}\ \bibnamefont {Stone}}, \
  and\ \bibinfo {author} {\bibfnamefont {A.}~\bibnamefont {Kohlmeyer}},\ }\href
  {\doibase 10.1016/j.jcp.2011.01.048} {\bibfield  {journal} {\bibinfo
  {journal} {Journal of computational physics}\ }\textbf {\bibinfo {volume}
  {230}},\ \bibinfo {pages} {3556} (\bibinfo {year} {2011})}\BibitemShut
  {NoStop}%
\bibitem [{\citenamefont {Ishizawa}\ \emph {et~al.}(1980)\citenamefont
  {Ishizawa}, \citenamefont {Miyata}, \citenamefont {Minato}, \citenamefont
  {Marumo},\ and\ \citenamefont {Iwai}}]{Ishizawa1980}%
  \BibitemOpen
  \bibfield  {author} {\bibinfo {author} {\bibfnamefont {N.}~\bibnamefont
  {Ishizawa}}, \bibinfo {author} {\bibfnamefont {T.}~\bibnamefont {Miyata}},
  \bibinfo {author} {\bibfnamefont {I.}~\bibnamefont {Minato}}, \bibinfo
  {author} {\bibfnamefont {F.}~\bibnamefont {Marumo}}, \ and\ \bibinfo {author}
  {\bibfnamefont {S.}~\bibnamefont {Iwai}},\ }\href {\doibase
  10.1107/S0567740880002981} {\bibfield  {journal} {\bibinfo  {journal} {Acta
  Crystallographica Section B Structural Crystallography and Crystal
  Chemistry}\ }\textbf {\bibinfo {volume} {36}},\ \bibinfo {pages} {228}
  (\bibinfo {year} {1980})}\BibitemShut {NoStop}%
\bibitem [{\citenamefont {El-mashri}\ \emph {et~al.}(1983)\citenamefont
  {El-mashri}, \citenamefont {Jones},\ and\ \citenamefont
  {Forty}}]{ElMashri1983}%
  \BibitemOpen
  \bibfield  {author} {\bibinfo {author} {\bibfnamefont {S.~M.}\ \bibnamefont
  {El-mashri}}, \bibinfo {author} {\bibfnamefont {R.~G.}\ \bibnamefont
  {Jones}}, \ and\ \bibinfo {author} {\bibfnamefont {A.~J.}\ \bibnamefont
  {Forty}},\ }\href {\doibase 10.1080/01418618308236536} {\bibfield  {journal}
  {\bibinfo  {journal} {Philosophical Magazine A}\ }\textbf {\bibinfo {volume}
  {48}},\ \bibinfo {pages} {665} (\bibinfo {year} {1983})}\BibitemShut
  {NoStop}%
\bibitem [{\citenamefont {Bourdillon}\ \emph {et~al.}(1984)\citenamefont
  {Bourdillon}, \citenamefont {El-mashri},\ and\ \citenamefont
  {Forty}}]{Bourdillon1984}%
  \BibitemOpen
  \bibfield  {author} {\bibinfo {author} {\bibfnamefont {a.~J.}\ \bibnamefont
  {Bourdillon}}, \bibinfo {author} {\bibfnamefont {S.~M.}\ \bibnamefont
  {El-mashri}}, \ and\ \bibinfo {author} {\bibfnamefont {a.~J.}\ \bibnamefont
  {Forty}},\ }\href {\doibase 10.1080/01418618408233278} {\bibfield  {journal}
  {\bibinfo  {journal} {Philosophical Magazine A}\ }\textbf {\bibinfo {volume}
  {49}},\ \bibinfo {pages} {341} (\bibinfo {year} {1984})}\BibitemShut
  {NoStop}%
\bibitem [{\citenamefont {Lee}\ \emph {et~al.}(2009)\citenamefont {Lee},
  \citenamefont {Lee}, \citenamefont {Park}, \citenamefont {Yi},\ and\
  \citenamefont {Ahn}}]{Lee2009}%
  \BibitemOpen
  \bibfield  {author} {\bibinfo {author} {\bibfnamefont {S.~K.}\ \bibnamefont
  {Lee}}, \bibinfo {author} {\bibfnamefont {S.~B.}\ \bibnamefont {Lee}},
  \bibinfo {author} {\bibfnamefont {S.~Y.}\ \bibnamefont {Park}}, \bibinfo
  {author} {\bibfnamefont {Y.~S.}\ \bibnamefont {Yi}}, \ and\ \bibinfo {author}
  {\bibfnamefont {C.~W.}\ \bibnamefont {Ahn}},\ }\href {\doibase
  10.1103/PhysRevLett.103.095501} {\bibfield  {journal} {\bibinfo  {journal}
  {Physical Review Letters}\ }\textbf {\bibinfo {volume} {103}},\ \bibinfo
  {pages} {4} (\bibinfo {year} {2009})}\BibitemShut {NoStop}%
\bibitem [{\citenamefont {Momida}\ \emph {et~al.}(2011)\citenamefont {Momida},
  \citenamefont {Nigo}, \citenamefont {Kido},\ and\ \citenamefont
  {Ohno}}]{Momida2011}%
  \BibitemOpen
  \bibfield  {author} {\bibinfo {author} {\bibfnamefont {H.}~\bibnamefont
  {Momida}}, \bibinfo {author} {\bibfnamefont {S.}~\bibnamefont {Nigo}},
  \bibinfo {author} {\bibfnamefont {G.}~\bibnamefont {Kido}}, \ and\ \bibinfo
  {author} {\bibfnamefont {T.}~\bibnamefont {Ohno}},\ }\href {\doibase
  10.1063/1.3548549} {\bibfield  {journal} {\bibinfo  {journal} {Applied
  Physics Letters}\ }\textbf {\bibinfo {volume} {98}},\ \bibinfo {pages}
  {042102} (\bibinfo {year} {2011})}\BibitemShut {NoStop}%
\end{thebibliography}%

\end{document}